\documentclass[12pt]{article}
\usepackage{latexsym,graphicx}
\newcommand{\be}{\begin{equation}}
\newcommand{\ee}{\end{equation}}
\usepackage[left=2.5cm,top=2.5cm,right=2.5cm,bottom=1.5cm]{geometry}
\usepackage{float}
\usepackage{epstopdf}
\usepackage{cite}
\begin{document}
\begin{center}
\large{\bf{Yukawa-Casimir wormholes in $f(Q)$ gravity
		}}\\
\vspace{10mm}
\normalsize{ Ambuj Kumar Mishra$^1$, Shweta$^2$,  Umesh Kumar Sharma$^3$}  \\
\vspace{5mm}
\normalsize{$^{1,2,3}$Department of Mathematics, Institute of Applied Sciences and Humanities, GLA University,
	Mathura-281 406, Uttar Pradesh, India}\\

\vspace{2mm}
$^1$E-mail:  ambuj\_math@rediffmail.com \\
\vspace{2mm}
$^2$E-mail:  shwetaibs84@gmail.com \\
 \vspace{2mm}
$^3$E-mail:  sharma.umesh@gla.ac.in \\
\vspace{2mm}
\end{center}
\vspace{10mm}
%\date{}
%\maketitle
\begin{abstract}
Casimir energy is always suggested as a possible source to create a traversable wormhole. It is also used to demonstrate the existence of negative energy, which can be created in a lab. To generalize, this idea, Yukawa modification of Casimir source has been considered in Remo Garattini (Eur. Phys. J. C 81 no.9, 824, 2021). In this work, we explore the  Yukawa Casimir wormholes in symmetric teleparallel gravity. We have taken four different forms of $f(Q)$ to obtain wormhole solutions powered by the original Casimir energy source and Yukawa modification of the Casimir energy source. In power law form $f(Q)= \alpha Q^2 + \beta$ and quadratic form $f(Q)= \alpha Q^2 + \beta Q + \gamma$, where $\alpha, \beta, \gamma$ are constants and $Q$ is non-metricity scalar, we analyze that wormhole throat is filled with non-exotic matter. We find self-sustained traversable wormholes in the Casimir source where null energy conditions are violated in all specific forms of $f(Q)$, while after Yukawa modification it is observed that violation of null energy conditions is restricted to some regions in the vicinity of the throat.
\end{abstract}



%%%%%%%%%%%%%%%%%%%%%%%%%%%%%%%%%%%%%%%%%%%%%%%%%%%% Section 1 %%%%%%%%%%%%%%%%%%%%%%%%%%%%%%%%%%%%%%%%%%%%%
\section{Introduction}
According to the well known quote of Wheeler, ``space tells matter how to move, matter tell space how to curve". Before Einstein the prevalent theory that ruled for over 200 years was the newton's theory which suggested that every object in space that has a mass, put a gravitational pull on each other. This theory was then discarded when Einstein gave the idea that the space-time is bend by the massive masses and this curvature of space results in gravity. It is evident that the most astounding theories in science are associated to the Einstein and the most revolutionary is the General theory of Relativity. With the emergence of the techniques like EHT, VIRGO, gravitational lenses it is experimentally proved to be eighty percent correct theory of gravity \cite{ref1,ref2,ref3,ref4}. During September 2015 to January 2016, advanced LIGO very first detected the gravitational waves from the merger of black-holes \cite{ref5} justifying another conjecture of GR that gravitational waves are result of the distortion in the curvature of space time\\   
  
  The solutions of the Einstein field equations define the GR further enabling to describe the other physical phenomena such as wormholes, black-holes, evolution of universe, inflation, planetary dynamics etc. The most sought after prediction of GR is the idea of traversable wormholes which provides almost instantaneous travel to arbitrarily distant locations in space and time. Wormholes are still hypothetical structures that are supposed to be tunnel like in shape which has two openings at each ends connected with a throat. Although the wormholes are completely hypothetical till now and there is no experimental proof of there existence, still the curiosity in them is exponentially increasing as they are assumed to be potential instrument for time travel and interstellar travel in affordable time span where affordable signifies to be within the human life time.\\

  The idea of static wormholes in the form of Schwarschild solutions, joining two asymptotically flat regions in space-time, was first given by Flamm \cite{ref6}. The solution of Einstein field equations coincide with the solution of traversable wormholes \cite{ref7}. The idea of traversable wormholes, also called Einstein Rosen bridges, was  first conceived by Einstein and Nathan Rosen\cite{ref8}. Eventually the solutions close to the Einstein Rosen bridge connecting two Schwarschild solutions were also achieved \cite{ref9,ref10}. Morrice and Thorne worked on the physical requirements to construct such apparatus and presented the idea of spherically symmetric traversable wormhole which subsequently proved to be in synchronization with tachyonic mass-less scalar field. Later on, Yurtsever along with them introduced the concept of time machine giving a possible solution of traversable wormhole \cite{ref11,ref12,ref13,ref14,ref15,ref16} .\\ 
  
  The GR is inconsistent with the present acceleration of universe \cite{ref16a}. Within the framework of GR only with the help of extremely fine-tuned cosmological or a typical dark energy source this phenomenon of universe  can be described \cite{ref16b}. This inconsistency escorts to the evolution of extensions of GR in the form of modified theories of gravity. The most simple theories among these extensions are the scalar tensor theories that can justify the accelerated expansion of universe \cite{ref16c,ref16d}. Secondly Nowadays researchers are more interested towards the solution of traversable wormholes free from any kind of singularity and any event horizon as the schwarschild metric shatter close to the event horizon. Hence some adjustments have to be made in the metric along with some restrictions on the wormhole throat by implementing Birkho theorem \cite{ref17}, in which case mass energy becomes less than the radial tension \cite{ref18,ref19,ref20,ref21} leading to the violation of NEC. This violation in-turn leads to the presence of exotic matter. To avoid this presence of exotic matter in space-time of wormhole, various modified theories of gravity were evolved where higher order curvature terms accounts for the violation of null energy condition. A thin shell traversable wormhole solution was produced in the framework of $f(R)$ gravity. The coupling of geometry with the matter terms gave rise to the  $f(R, T)$ theory of gravity and several wormhole solutions are studied for several forms of $f(R, T)$  \cite{ref22,ref23,ref24,ref25, ref25a,ref26,ref27,ref28,ref28a,ref28b, ref29,ref29a,ref29b,ref30}.\\
 
  Teleparallel and other modified theories of gravity also played a major role in exploring the solutions of traversable wormholes \cite{ref31,ref32,ref33,ref34,ref35,ref35a}. Teleparallel theory \cite{ref36,ref37} has been proved to be one of the potential alternatives of GR. Here the geometry is free from torsion and curvature while the gravitational interactions of the space-time are portrayed by the non-metricity term $Q$. The Lagrangian of $f(Q)$ gravity is reformed in explicit function of red-shift, $f(z)$ to justify these teleparallel gravity models by means of various observational tools like Gamma Ray Bursts, Cosmic Microwave Background, Baryon Acoustic Oscillations data, quasars etc \cite{ref38,ref39}. Various researchers established different gravitational modification classes implementing different function forms of $f(Q)$ and discussed the wormhole solutions and other cosmographical results \cite{ref40,ref41,ref42,ref43,ref44,ref45,ref46,ref47,ref47a,ref47b, ref47c}.\\
  
  As earlier stated an unfortunate consequence in GR is the unavoidable association of traversability of wormhole with the violation of NEC which implies towards the exotic matter treading the wormhole throat which has negative energy and here derives our motivation of present work as we are focused to explore the solution of traversable wormhole space-time threaded by the ordinary matter. We try to find such solution where the negative energy required to hold open the throat for requisite time can be sourced from some other source of negative energy. As the classical matter justifies the null energy condition the wormhole solution can be obtained in the framework of semi classical or quantum field. In that case Casimir energy density of Casimir device could be an appropriate source of exotic energy as till date Casimir energy is the only artificial source of negative energy \cite{ref48}. The Casimir energy arise in vacuum between two closely placed, uncharged, plane parallel metallic plates. an attractive force appears in this event which was predicted in 1948 and later on confirmed experimentally in the Philips Laboratories \cite{ref49,ref50} and by other investigators in recent years \cite{ref51,ref52,ref52b}.\\
   
  A modification to the original Casimir shape function \cite{ref52a} is made by implementing Yukawa term. The zero tidal force condition (ZTF) was imposed by modifying the original profile with Yukawa type modifications to procure different solutions to get the signals of traversable wormholes and to obtain the possibility of getting negative energy density more concentrated in the proximity of throat, as they stated in their article \cite{refc}.
   
   In the present work we investigate the probable solutions of traversable wormholes with ordinary matter devising the Yukawa-Casimir shape function in the backdrop of symmetric teleparallel gravity where four different functional forms of $f(Q)$ are discussed. First case showcases the linear form. Second and third cases analyze Yukawa-Casimir wormhole solution for power law forms. Fourth case is the inverse power law form of teleparallel gravity.

%%%%%%%%%%%%%%%%%%%%%%%%%%%%%%%%%%%%%%%%%%%%%%%%%%%% Section 2done %%%%%%%%%%%%%%%%%%%%%%%%%%%%%%%%%%%%%%%%%%%%%%%%
\section{Symmetric Teleparallel Gravity i.e $f(Q)$-Gravity} 
 The action corresponding to the symmetric teleparallel gravity, taken in this article is \cite{ref53}

\begin{equation}\label{eq1}
	S = \frac{1}{2} \int \left[ f\left(Q \right) + {2\mathcal{L}_{\mu}} \right] \sqrt{-g}  d^4 x,  
\end{equation}
Here, the function $f(Q)$ regards to the non-metricity term Q whereas the matter lagrangian density can be described as $ \mathcal{L}_{\mu}$ and the determinant of the metric  ${g}_{xy}$ is given as $g$.
 The equation for non-metricity tensor is written as

\begin{equation}\label{eq2}
	{Q}_{\lambda xy} = \nabla _{\lambda xy},  
\end{equation}
 The two independent traces corresponding to non-metricity tensor come out to be 

\begin{equation}\label{eq3}
	{Q}_{\phi} = {{{Q}_{\phi}} ^ {x}} _ {x}  \qquad  , \qquad  \bar{Q}_{\phi} = {{Q}^{x}} {_ { \phi x}}
\end{equation}
 
Jiminez et.al. \cite{ref37} addressed this new geometry as new GR. Also they defined the non-metricity conjugate related to the super-potential of this so called new GR as 

\begin{equation}\label{eq4}
	P^\phi_{xy} = \frac{1}{4} \left[- {Q^\phi}_{xy} + 2 {Q_{xy}^\phi }  + {Q^\phi} g_{xy} - \bar{Q}^{\phi} g_{xy}- \delta^\phi_{(xQ_y)} \right]
\end{equation}
which can be snatched by contemplating the following form of non-metricity tensor

\begin{equation}\label{eq5}
	Q =  - Q _ {\phi xy} P ^{\phi xy} ,
\end{equation}
The features of matter threading the space-time are defined by the energy momentum tensor which is stated as

\begin{equation}\label{eq6}
	T _ {xy} = - \frac{2}{\sqrt{-g}} \frac{ \delta \left(\sqrt{-g} \mathcal{L}_{\mu} \right)}{ \delta g^{xy}},
\end{equation} 

We get the motion equations by varying the action \ref{eq1} with respect to the metric tensor $g _ {xy}$ as

\begin{equation}\label{eq7}
	\frac{2 \nabla _ \eta}{\sqrt{- g }} \left( \sqrt {-g} {f_Q} {P^\eta} _{xy} \right) + \frac{1}{2} g_{xy} f + {f_Q} \left( P _{x \eta \zeta} {{Q_y} ^{\eta \zeta}} - 2 {Q}_{\eta \zeta x}  {{P^\eta \zeta} _ y} \right) = - {T} _{xy},
\end{equation} 

Here $ {f}_{Q}$ represents the total derivative of $f$ with respect to $Q$. The variation of Eq.(\ref{eq1}) regarding the connections gives the following relation

\begin{equation}\label{eq8}
	{\nabla _ x}{\nabla _ y} \left( \sqrt{-g} {f_Q} {P^\eta} _{xy} \right) = 0
\end{equation} 

The field equations assert to conserve the energy momentum tensor with the conformity of specified $f(Q)$ gravity. In the present study we focus to specify the gravitational field equations commanding the static and spherically symmetric solutions \cite{ref54} to the wormhole geometry.

%%%%%%%%%%%%%%%%%%%%%%%%%%%%%%%%%%%%%%%%%%%%%%%%%%%%%%%%%  done below
\section{ Wormhole Geometry and Solution of Field Equations in $f(Q)$ Gravity}

The usual spherically symmetric and static line element of Morris-Thorne class to specify the wormhole geometry is written as
\begin{equation}\label{eq9}
	ds^2= \exp \left(2 \phi(r)\right) dt^2 + {\left ( \frac{r - b \left( r \right)}{ r} \right) ^{-1}} d r^2 +{r^2} d {\theta}^2 + {r^2} {\sin}^2 {\theta} d{\phi}^2
\end{equation}
Here, the redshift function of the intervening object with regards to the radial co-ordinate $r$ is given by $\phi (r)$. By the definition of radial co-ordinate $r$ as $ 0 < r_0 \leq r \leq \infty$, its non-monotonic behavior is indicated. As one can easily collect that $r$ falls from $\infty$ and approaches  $b(r_0)$ which is the minimum value $r_0$ and after attaining this minimum  value it again proceeds to infinity. This minimum $r_0$ is the throat radius. To comply with the traversability of wormhole it has to be prevented from any event horizon or the presence of singularity. The redshift function therefore is restricted to attain only a non-zero finite value to avoid such occurrence. In our present article we are examining the Yukawa-Casimir wormhole \cite{refc} solutions for which the shape function and redshift functions are fixed and are obtained by introducing Yukawa terms in original Casimir wormhole \cite{ref55}. Therefore the shape function also satisfies all the required constraints such as flaring out condition, throat condition, asymptotically flatness condition etc.

The matter fluid that threads the wormhole throat is considered as anisotropic which has the stress-energy-momentum tensor

	\begin{equation}\label{eq10}
		T_x ^y = \left(\rho + p_t\right) {u_x} {u^y} - {p_t}{\delta_x ^y} + \left(p_r - p_t \right){v_x} {v^y} ,
	\end{equation} 
	 
	 Here, $v_x$ and $u_x$ are unitary space-like vector in radial direction and four velocity respectively, $\rho$, $p_r$, $p_t$ represent the energy density and principal pressures respectively.
	 
	 The trace $Q$ of non-metricity tensor, for the line element Eq.(\ref{eq9}), within $F(Q)$  gravity is given by
	
	\begin{equation}\label{eq11}
		Q = - {\frac{2}{r^3}} \left[ r - b(r) \right] \left[ 2.r.{\phi'(r)} + 1\right] ,
	\end{equation} 
	
	We solve Eq.(\ref{eq7}), Eq.(\ref{eq9}) and Eq.(\ref{eq10}) to get the expressions for energy density ($\rho$), radial pressure ($p_r$), tangential pressure ($p_t$) in terms of radial coordinate $r$ as

	\begin{equation}\label{eq12}
		\rho =	\left[ {\frac{1}{r^3}} \left( 2 r \left( r - b(r) \phi '(r)- r b'(r) - b(r) +{r}   \right) \right) \right] f_Q - \frac{2}{r^2}\left( b(r)-r \right) f_Q + \frac{f}{2}, 
		\end{equation} 
		
		\begin{equation}\label{eq13}
			p_r = - \left[ {\frac{2}{r^3}} \left(r - b(r)\right) \left( 2 r \phi'(r) +1\right) -1 \right] f_Q -\frac{f}{2} ,
		\end{equation} 
			\begin{eqnarray}\label{eq14}
			p_t &=& - \left[ {\frac{1}{r^3}}\left( \left[r^2 \phi''(r)+ r \phi'(r)  \left(r \phi'(r) +3 \right) +1\right] \left[r - b(r)\right]\right.\right. \nonumber\\
			 &-&
			 \left.\left.\frac{1}{2}\left[  r \phi'(r)+ 1\right]\left[r b'(r) - b(r) \right] \right)  \right]f_Q - \frac{1}{r^2}\left[ r \phi'(r)+1\right] \left[r - b(r)\right] f_Q - \frac{f}{2},
		\end{eqnarray} 
		
		With appropriate choices of the shape function $b(r)$ and the redshift function $\phi (r)$, one can analyze the plausible solutions for wormhole geometry in the backdrop of modified gravity. In this document we take the specified Yukawa-Casimir shape function and redshift function to acknowledge the negative energy in the wormhole geometry.

%%%%%%%%%%%%%%%%%%%%%%%%%%%%%%%%%%%%%%%%%%%%%%%%%%%%%%%%%%%%%%%%%%%%%%%%%%%

%%%%%%%%%%%%%%%%%%%%%%%%%%%%%%%%%%%%%%%%%%%%%%%%%%%%%%%%%Done
\subsection{The Energy conditions}
 The exotic properties of the matter field are the necessary requirements of wormhole space-time. The Energy conditions are the tool that play the prime role in defining these exotic properties. These energy conditions are particular constraints made on the matter stress-energy-momentum  tensor which demonstrates the basic features of various forms of matter. The Idea of energy conditions and their various implications are summarized in \cite{ref55a}. These constraints are extracted from the Raicaudhary equations that narrates the temporal evolution of expansion scalar $\theta$ for the congruences of the time-like vector  $u^l$ and $k_l$ which in turn describes the null geodesics as
 
 \begin{equation}\label{15}
 	\frac{d\theta}{d\tau} - \omega _{lm}\omega^{lm} + \sigma_{lm}\sigma^{lm} + \frac{1}{3} \theta^2 + R_{lm} u^l u_m = 0
 \end{equation}
 
 \begin{equation}\label{16}
 	\frac{d\theta}{d\tau} - \omega _{lm}\omega^{lm} + \sigma_{lm}\sigma^{lm} + \frac{1}{2} \theta^2 + R_{lm} k^l k_m = 0
 \end{equation}
 
 where $k^l$ denotes the vector field, on the other hand, the shear or spatial tensor is given by  $R_{lm} k^l k_m$ having  $\sigma^{2}= \sigma_{lm}\sigma^{lm} \geq 0$ and $\omega_{lm}\equiv0$.
 
 These conditions are used to get an idea about the basic features of fluid threading the wormhole throat. The energy conditions are normally examined with regard to principle pressures and energy density. The justification or infringement of energy conditions are responsible for the occurrence, existence and stability of the traversable wormholes. These conditions coincide with the Raichaudhary conditions for $\theta<0$ i.e. for attractive geometry or in other words, for positive energy.
 
 	\begin{equation}\label{17}
 	R_{l m} u^l u_m \geq 0
 \end{equation}
 
 \begin{equation}\label{18} 
 	R_{l m} k^l k_m \geq 0
 \end{equation}
 
 Here, the energy conditions are discussed for the anisotropic matter fluid. The enegy conditions can be expressed in terms of energy density $\rho$, radial pressure $p_r$ and tangential pressure $p_t$ as  $\forall i,\, \rho(r) + p_{i}\geq 0$. The tensor form of NEC is given as  $T_{lm} k^{l}k^{m} \geq 0$. The NEC implies the non-negativity of the principle pressures. The traversability of WH is unfortunately consequent to the violation of the NEC in GR. The weak energy condition or WEC assures that the energy density of a time like vector can not be negative. With regard to principle pressures, WEC is given as $\rho(r)\geq 0$ and $\forall i,\,\, \rho(r) + p_{i}\geq 0$ while the tensor form of WEC is $T_{lm} k^{l} k^{m} \geq 0$. $(T_{lm}-\frac{T}{2}g_{lm}) k^{l}k^{m} \geq 0 $ is the tensor form of strong energy condition whereas in principle pressures SEC can be written as $ T = -\rho(r) + \sum_{j} p_{j}$ and  $\forall j,\,\, \rho(r) + p_{j}\geq 0,\,\,  \rho(r) +\sum_{j} p_{j}\geq 0 $ whose violation is mandatory for the inflation of the universe. The Dominant energy condition is the one which puts a restriction on the transmission of energy and limits its rate to to speed of light. Both tensor and principle pressure forms of DEC are expressed as $ T_{lm} k^{l}k^{m} \geq 0$ and $\rho(r) \geq 0$ and $\forall i,\,\, \rho \pm p_{i} \geq0$ respectively. Here $T_{lm}k^{l}$ is not space-like.

%%%%%%%%%%%%%%%%%%%%%%%%%%%%%%%%%%%%%%%%%%%%%%%%%%%%%%%% Section 4 %%%%%%%%%%%%%%%%%%
\section{The Yukawa-Casimir Wormhole Model} 

In this article we examine the Yukawa-Casimir wormhole solutions in the backdrop of four different form functions of $f(Q)$ representing the symmetric teleparallel gravity. For Yukawa-Casimir wormhole model it is assumed that the exotic matter is substituted by the Casimir energy density. Thus far Casimir energy is the only known artificial negative energy source which has the energy density $\rho_0 = - \frac{hc\pi^2}{720 d^4}$.

The stress energy tensor (SET) of Casimir energy is 

\begin{equation}
	T_{\mu\nu} = \frac{hc \pi^2}{720 d^4}\left[dia\left(-1,-3,1,1\right)\right]
\end{equation}

here $d$ represents the plate separation which is traceless and whose divergence is also null. By establishing the connection between above SET and the space-time metric, the Casimir shape function $b(r)$ and redshift function $\phi(r)$  are obtained as \cite{ref52a}

\begin{equation}\label{18}
	b(r) = \frac{2 r_0}{3} + \frac{{r_0}^2}{3r} ,  \quad \phi(r) = \ln \left(\frac{3r}{3r +r_0}\right)
\end{equation}

 The Casimir wormhole described by above shape function does not satisfy the zero tidal force (ZTF) condition. In \cite{refc} authors deformed the Casimir wormhole shape function by inducing the Yukawa profile of the form ${\exp\left(-\mu\left(r - r_0\right)\right)}$ which also satisfy the ZTF condition. They found a possibility of a new family of solutions having vanishing redshift function. The Yukawa-Casimire shape function is thus obtained as 

\begin{equation}\label{19}
	b(r) = \frac{2 r_0}{3} + \frac{{r_0}^2}{3r}{\exp\left(-\mu\left(r - r_0\right)\right)} , 
\end{equation}
 
 here $\mu$ is a positive mass scale and the original Casimir profile is achieved again by putting $\mu=0$.

\subsection{Linear form:  $f(Q)=\alpha Q$}

The linear functional form $f(Q)=\alpha Q$ is considered here to analyze the Yukava-Casimir wormhole solution. This particular form helps in comparing the wormhole solutions with the usual wormholes and identifies with the symmetric teleparallel equivalent of general relativity. We devise the Yukawa Casimir system in the equations (\ref{eq12}),(\ref{eq13}) and (\ref{eq14}) to obtain the Yukawa-Casimir energy density and principal pressures as

\begin{eqnarray}\label{eq20}
\rho& =&\frac{1}{3 r^4 (3 r+{r_0})}\left\lbrace\alpha  e^{\mu  (-r)} \left(6 r^2 e^{\mu  r} \left(3 r^2+r ({r_0}-3)-3 {r_0}\right)-{r_0} e^{\mu  {r_0}} \left(6 (\mu +2) r^3\right.\right.\right.\nonumber\\
&+&
\left. \left.\left. r^2 (5 (\mu +2) {r_0}-12)+r {r_0} ((\mu +2) {r_0}-15)-5 {r_0}^2\right)\right)\right\rbrace
\end{eqnarray}  

\begin{eqnarray}\label{eq21}
p_r&=& \frac{1}{r^4 (3 r+{r_0})}\left\lbrace\alpha  {r_0} e^{\mu  ({r_0}-r)} \left(-2 r^2 \left(e^{\mu  (r-{r_0})}-1\right)+3 r {r_0}+{r_0}^2\right)\right\rbrace
\end{eqnarray}

\begin{eqnarray}\label{eq22}
p_t&=& -\frac{1}{6 r^4 (3 r+{r_0})^2}\left\lbrace\alpha  e^{\mu  (-r)} \left(6 r^2 e^{\mu  r} \left(9 r^3+9 r^2 {r_0}+r {r_0} (2 {r_0}-3)+{r_0}^2\right)\right.\right.\nonumber\\
&+&
\left. \left.{r_0} e^{\mu  {r_0}} \left(18 \mu  r^9+9 r^8 (3 \mu  {r_0}+2)+r^7 {r_0} (13 \mu  {r_0}+36)+2 r^6 {r_0}^2 (\mu  {r_0}+11)\right.\right.\right.\nonumber\\
&+&
\left.\left.\left. 4 r^5 {r_0}^3-36 r^4-54 r^3 {r_0}+2 r^2 {r_0} (6-13 {r_0})+2 r {r_0}^2 (1-2 {r_0})-2 {r_0}^3\right)\right)\right\rbrace
\end{eqnarray}

 The energy density $\rho$ and energy conditions regarding the Casimir energy source i.e. for $\mu =0$ and Yukawa-Casimir energy source i.e. for $\mu= 1,2,3$  are plotted against the radial coordinate $r$ and throat radius $r_0$ in Fig. 1 to Fig. 3. To investigate the Yukawa-Casimir wormhole geometry, we analyze the implications of these figures. From Fig. 1(a), it is evident that energy density $\rho$ is non-negative for all values $\mu=0, 1, 2, 3$ in the region $r\geq 1.6$. The radial and tangential NECs are mapped in Fig. 1(b) and 2(a) which shows that $\rho +p_r \geq 0$  for $r \in (1.6, \infty)$ while  $\rho +p_l < 0$  for $r\geq r_0$ for $\mu = 0, 1, 2$ but only for $\mu=3$ the tangential NEC is also validated for radial parameter $r\in (3.5, \infty)$. Therefore it is gathered that both the Null energy conditions are satisfied only for the particular value $\mu=3$ for the region $r > 3.5$ which indicates that Yukawa modification in the Casimir source is useful to minimize the exotic matter in wormhole formation. For this particular case, a solution with ordinary matter threading the throat is obtained and it can be the Casimir energy that is providing the required exotic or negative energy to sustain the wormhole. For other cases, even in the case of $\mu=0$, when the system reduces to the Casimir wormhole, the NECs are violated and this violation indicates the existence of exotic matter near the throat which supports the wormhole geometry. The SEC and DECs are also plotted in Figs. 2(b), 3(a), and 3(b), which are also violated.

%%%%%%%%%%%%%%%%%%%%%%%%%%%%%%%%%%%%%%%%%%%%%%%%%%%%%%%
%%%%%%%%%%%%%%%%%%%%%%%%%%%%%%%%%%%%%%%%%%%%%%%%%%%%%%%%%%
 %%%%%%%%%%%%%%%%%%%%%%%%%%%%%%%%%%%%%%%%%%%%%%%%%%%
 
 %%%%%%%%%%%%%%%%%%%%%%%%%%%%%%%%%%%%%%%%%%%%% Figure 2 %%%%%%%%%%%%%%%%%%%%%%%%%%%%%%%%%%%%%%%%%%%%%%%%%%%%%%%%%%%%
 \begin{figure}
 	(a)\includegraphics[width=7cm, height=6cm, angle=0]{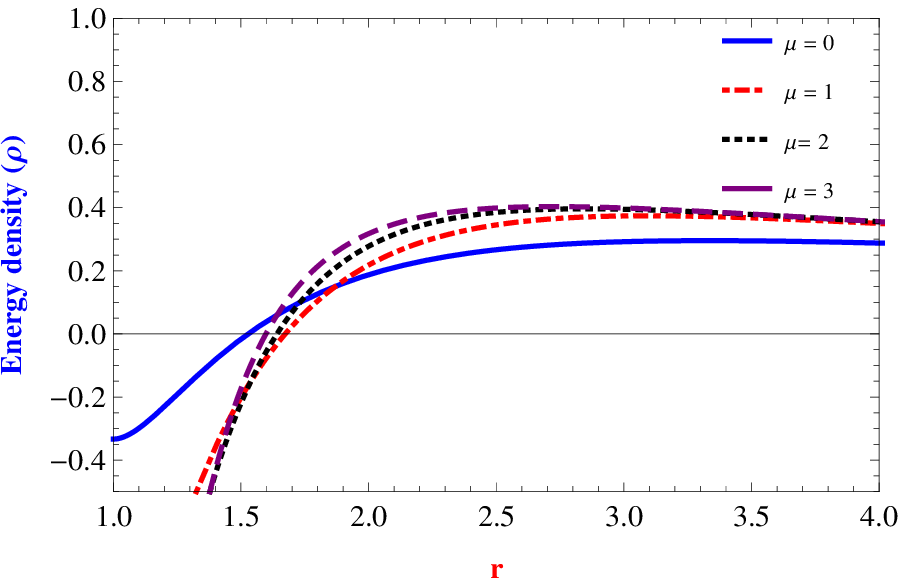}
 	(b)\includegraphics[width=7cm, height=6cm, angle=0]{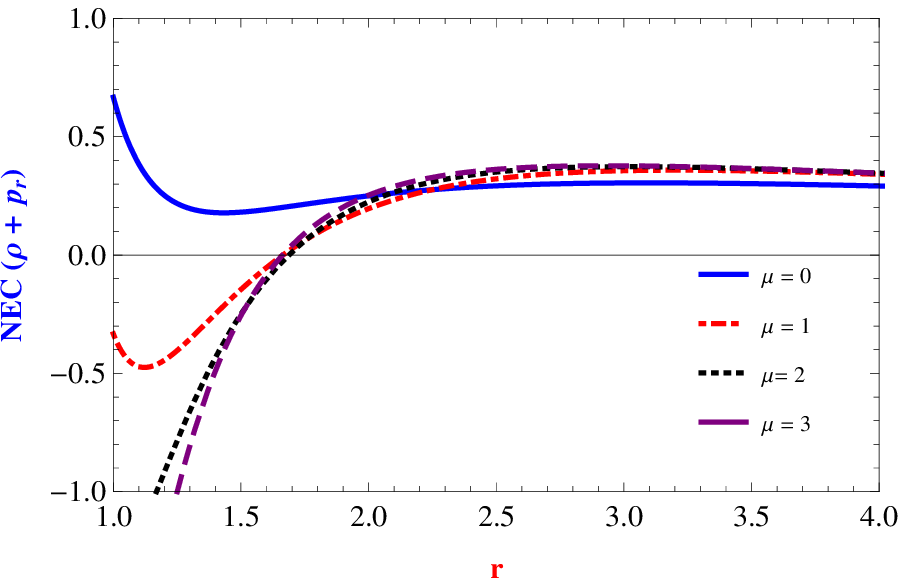}
 	\caption {Plots of energy density ($\rho$) and radial NEC ($\rho + p_r$) with throat radius $r_0 =1$ in $f(Q)= \alpha Q$ gravity}.
 \end{figure}
 %%%%%%%%%%%%%%%%%%%%%%%%%%%%%%%%% Figure 3  %%%%%%%%%%%%%%%%%%%%%%%%%%%%%%%%%%%%%%%%%%%%%%%%%%%%%%%%%%%%%%%%%%%%%%%%%%%%%%
 
 \begin{figure}
 	(a)\includegraphics[width=7cm, height=6cm, angle=0]{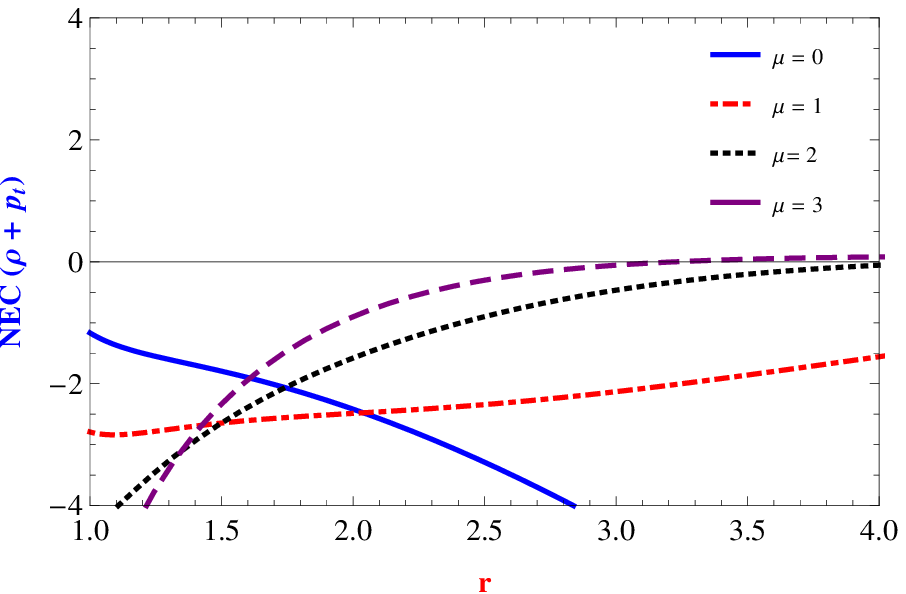}
 	(b)\includegraphics[width=7cm, height=6cm, angle=0]{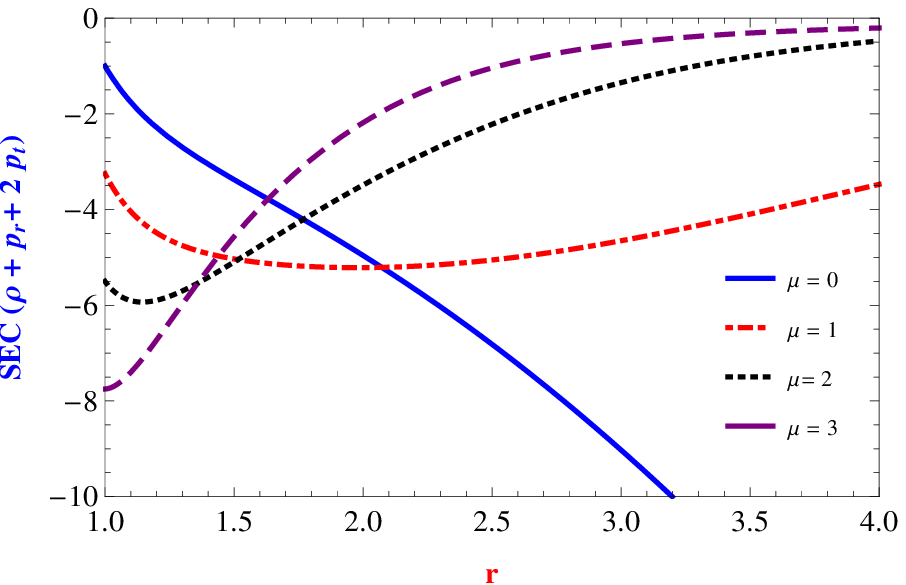}
 	\caption {Plots of tangential NEC ($\rho +p_t$) and SEC ($\rho + p_r+ 2p_t$) with throat radius $r_0=1$ in $f(Q)= \alpha Q$ gravity.}		
 	
 \end{figure}
 %%%%%%%%%%%%%%%%%%%%%%%%%%%%%%%%%%%%%%%%%%%%%% Figure 4 %%%%%%%%%%%%%%%%%%%%%%%%%%%%%%%%%%%%%%%%%%%%%%%%%%%
 \begin{figure}
 	(a)\includegraphics[width=7cm, height=6cm, angle=0]{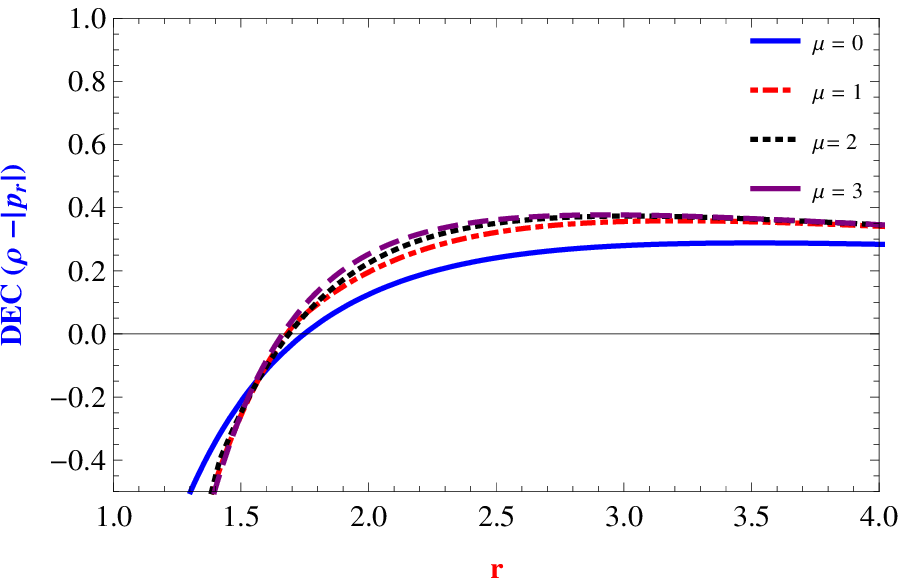}
 	(b)\includegraphics[width=7cm, height=6cm, angle=0]{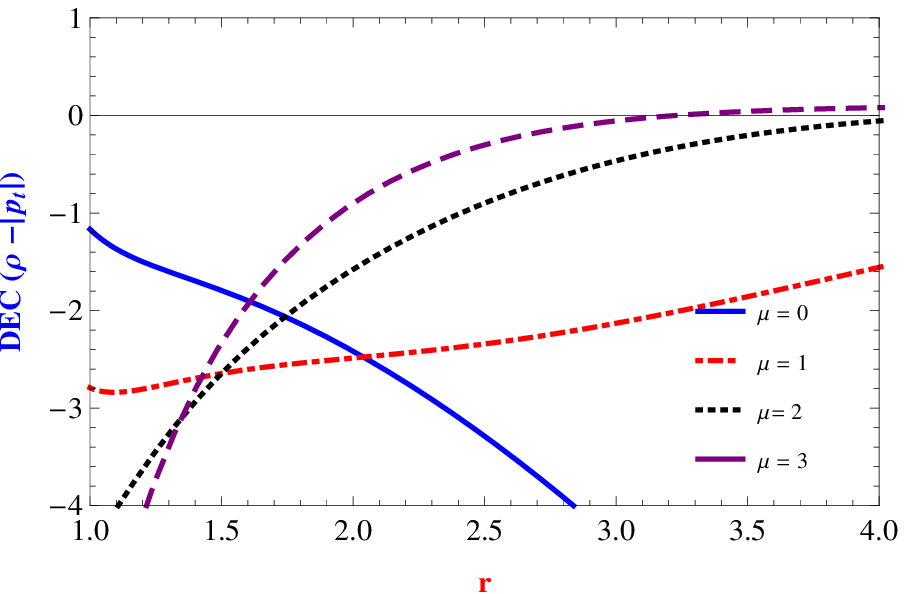}
 	\caption {Plots of radial  DEC ($\rho -|p_r|)$, and tangential DEC ($ \rho-|p_t|$) with throat radius $r_0 = 1$ in $f(Q)= \alpha Q$ gravity.}
 \end{figure}
 
 %%%%%%%%%%%%%%%%%%%%%%%%%%%%%%%%%%%%%%%%%%%%%%%%%%%%%%%%%%%% %%%%%%%%%%%%%%%%%%%%Figure5%%%%%%%%%%%%%%% 

\subsection{Power Law Form:  $f(Q)=\alpha Q^2 + \beta$}

As we saw in the above linear case that only for a particular case the traversable wormhole geometry filled with ordinary matter is found, we further proceed to investigate the Yukawa-Casimir wormhole solution with ordinary matter in the framework of power law form $f(Q)=\alpha Q^2 + \beta$.
 The radiation and CDM dominated background has already been acknowledged by this power law form. For this case the energy components are obtained by solving the FE, as given below

\begin{eqnarray}\label{eq23}
\rho &=&\frac{1}{6 r^8 (3 r+{r_0})^2}\left\lbrace e^{2 \mu  ({r_0}-r)} \left(24 \alpha  r^2 {r_0} (r+{r_0}) e^{\mu  (r-{r_0})} \left(6 (\mu +4) r^3+r^2 (5 (\mu +4) {r_0}-18)\right.\right.\right.\nonumber\\
&+&
\left. \left.\left. r {r_0} ((\mu +4) {r_0}-24)-8 {r_0}^2\right)-4 \alpha  {r_0}^2 \left(2 r^2+3 r {r_0}+{r_0}^2\right) \left(12 (\mu +2) r^3 \right.\right.\right.\nonumber\\
&+&
\left. \left.\left. 2 r^2 (5 (\mu +2) {r_0}-9)+r {r_0} (2 (\mu +2) {r_0}-21)-7 {r_0}^2\right)+3 r^4 e^{2 \mu  (r-{r_0})} \left(9 \beta  r^6 \right.\right.\right.\nonumber\\
&+&
\left. \left.\left. 6 \beta  r^5 {r_0}+\beta  r^4 {r_0}^2-144 \alpha  r^3-12 \alpha  r^2 (16 {r_0}-9)-24 \alpha  r {r_0} (2 {r_0}-9)+108 \alpha  {r_0}^2\right)\right)\right\rbrace
\end{eqnarray}

\begin{eqnarray}\label{eq24}
p_r& =&\frac{1}{2 r^8 (3 r+{r_0})^2} \left\lbrace e^{2 \mu  ({r_0}-r)} \left(12 \alpha  {r_0}^2 \left(2 r^2+3 r {r_0}+{r_0}^2\right)^2-16 \alpha  r^2 {r_0} \left(6 r^3+17 r^2 {r_0}\right.\right.\right.\nonumber\\
&+&
\left. \left.\left. 15 r {r_0}^2+4 {r_0}^3\right) e^{\mu  (r-{r_0})}+r^4 \left(-e^{2 \mu  (r-{r_0})}\right) \left(9 \beta  r^6+6 \beta  r^5 {r_0}+\beta  r^4 {r_0}^2-36 \alpha  r^2 \right.\right.\right.\nonumber\\
&-&
\left. \left.\left. 120 \alpha  r {r_0}-84 \alpha  {r_0}^2\right)\right)\right\rbrace
\end{eqnarray}

\begin{eqnarray}\label{eq25}
p_t &=&\frac{1}{6 r^8 (3 r+{r_0})^3} \left\lbrace e^{2 \mu  ({r_0}-r)} \left(-3 r^4 e^{2 \mu  (r-{r_0})} \left(27 \beta  r^7+27 \beta  r^6 {r_0}+9 \beta  r^5 {r_0}^2 \right.\right.\right.\nonumber\\
&+&
\left. \left.\left. r^4 \left(\beta  {r_0}^3-216 \alpha \right)-108 r^3 (\alpha +4 \alpha  {r_0})-12 \alpha  r^2 {r_0} (22 {r_0}+15)-12 \alpha  r {r_0}^2 (4 {r_0}+11) \right.\right.\right.\nonumber\\
&-&
\left. \left.\left. 60 \alpha  {r_0}^3\right)+12 \alpha  r^2 {r_0} (r+{r_0}) e^{\mu  (r-{r_0})} \left(18 \mu  r^9+9 r^8 (3 \mu  {r_0}+2)+r^7 {r_0} (13 \mu  {r_0}+36) \right.\right.\right.\nonumber\\
&+&
\left. \left.\left. 2 r^6 {r_0}^2 (\mu  {r_0}+11)+4 r^5 {r_0}^3-72 r^4-36 r^3 (3 {r_0}+1)-2 r^2 {r_0} (26 {r_0}+21)  \right.\right.\right.\nonumber\\
&-&
\left. \left.\left.  8 r {r_0}^2 ({r_0}+4)-10 {r_0}^3\right)-4 \alpha  {r_0}^2 \left(2 r^2+3 r {r_0}+{r_0}^2\right) \left(18 \mu  r^9+9 r^8 (3 \mu  {r_0}+2)  \right.\right.\right.\nonumber\\
&+&
\left. \left.\left. r^7 {r_0} (13 \mu  {r_0}+36)+2 r^6 {r_0}^2 (\mu  {r_0}+11)+4 r^5 {r_0}^3-36 r^4-18 r^3 (3 {r_0}+1) \right.\right.\right.\nonumber\\
&-&
\left. \left.\left. r^2 {r_0} (26 {r_0}+21)-4 r {r_0}^2 ({r_0}+4)-5 {r_0}^3\right)\right)\right\rbrace 
\end{eqnarray}

The energy density $\rho$ and the energy conditions for this case are mapped again for $\mu=0,1,2,3$. The energy density for this case, as depicted in Fig. 4(a) is positive for all four values of $\mu$ and $\forall r\geq r_0$. Here,
we are concerned about the solutions which are supported by the ordinary matter, which is achieved by the non-violation of NECs. From Fig. 4(b) we found the radial NEC is satisfied for the Yukawa-Casimir energy source at the throat and in $r\in (1, 1.65)$. It is interesting to see that radial NEC is violated in the case of Casimir energy for $\forall r\geq r_0$. From Fig. 5(a) one can clearly see that the tangential NEC is satisfied throughout the region for every value of $\mu$ including the case of the Casimir wormhole where $\mu=0$ which signifies that except for the case of $\mu=0$ the wormhole solutions with ordinary matter near the throat are obtained. The SEC is plotted in Fig. 5(b), which is violated at the throat in each case. The DECs are also shown in Figs. 6(a) and 6(b) which is also satisfied at the throat for $\mu = 1,2,3$ and radial DEC is violating for $\mu =0$.   

 %%%%%%%%%%%%%%%%%%%%%%%%%%%%%%%%%%%%%%%%%%%%%%%%%%%%%%%%%%%%%%
 %%%%%%%%%%%%%%%%%%%%%%%%%%%%%%%%%%%%%%%%%%%%%%%%%%%%%%%%%%%%%%%%%%
 
  %%%%%%%%%%%%%%%%%%%%%%%%%%%%Figure6%%%%%%%%%%%%%%%%%%%%%%%%%%%%%%%%
  %%%%%%%%%%%%%%%%%%%%%%%%%%%%%%%%%%%%%%%%%%%%%%%%%%%%%%%%%%%
  
  \begin{figure}
  	(a)\includegraphics[width=7cm, height=6cm, angle=0]{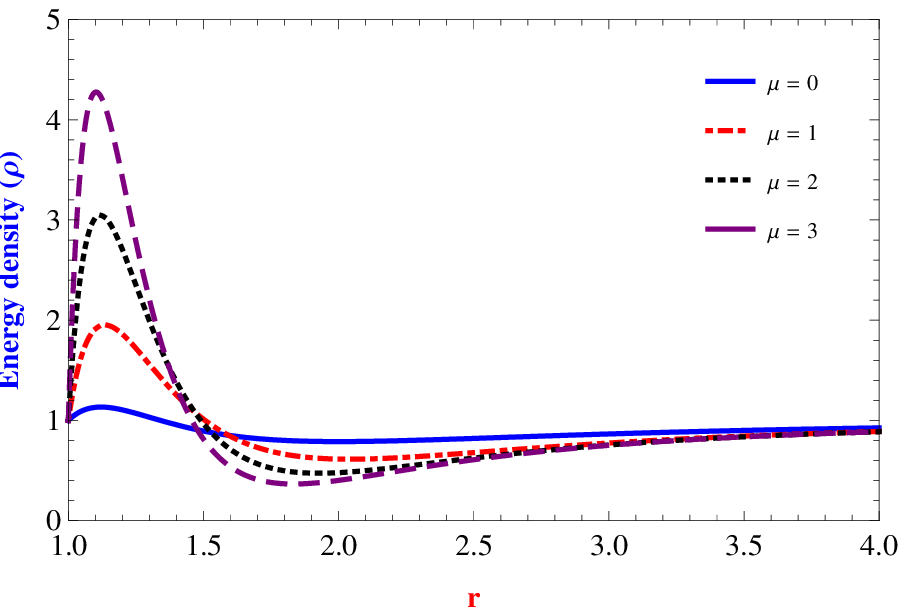}
  	(b)\includegraphics[width=7cm, height=6cm, angle=0]{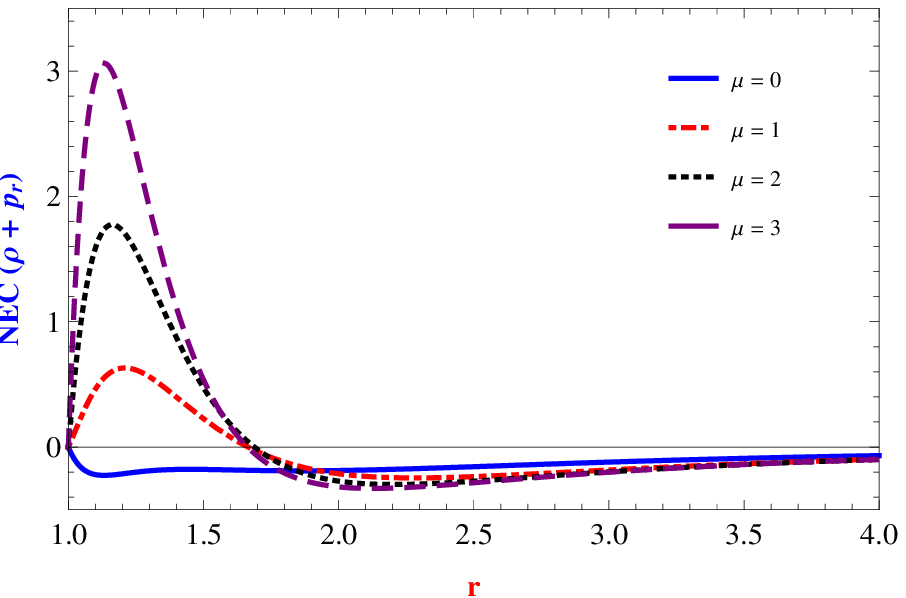}
  	\caption {Plots of energy density ($\rho$) and radial NEC ($\rho + p_r$) with throat radius $r_0 =1$ in  $f(Q)=\alpha Q^2 + \beta$ gravity}.
  \end{figure}
  %%%%%%%%%%%%%%%%%%%%%%%%%%%%%%%%% Figure 7 %%%%%%%%%%%%%%%%%%%%%%%%%%%%%%%%%%%%%%%%%%%%%%%%%%%%%%%%%%%%%%%%%%%%%%%%%%%%%%%
  \begin{figure}
  	(a)\includegraphics[width=7cm, height=6cm, angle=0]{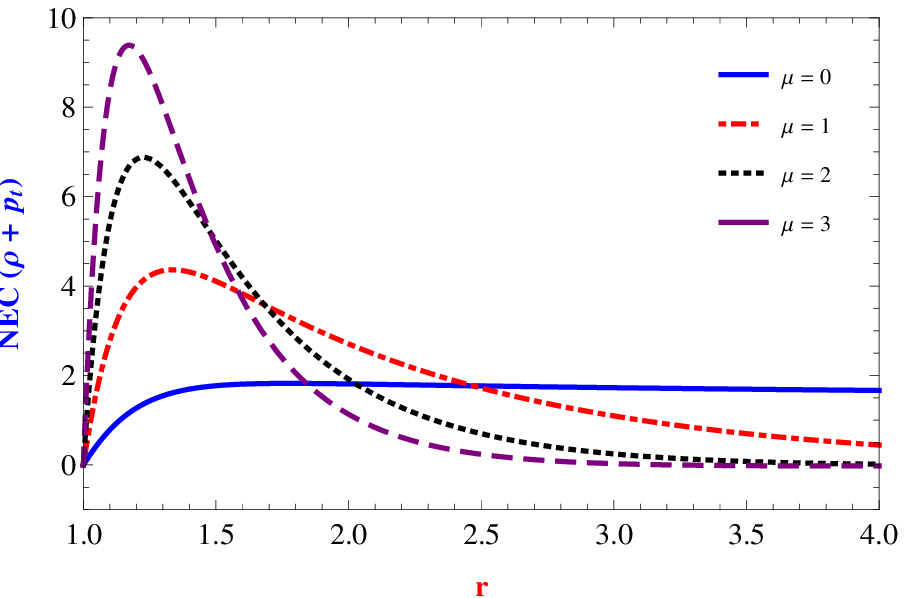}
  	(b)\includegraphics[width=7cm, height=6cm, angle=0]{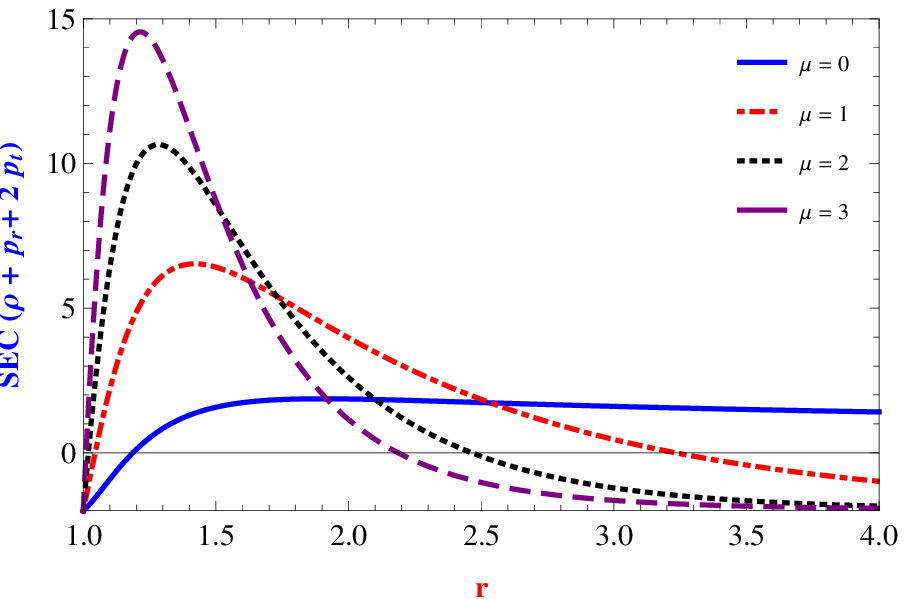}
  		\caption {Plots of tangential NEC ($\rho +p_t$) and SEC ($\rho + p_r+ 2p_t$) with throat radius $r_0=1$ in  $f(Q)=\alpha Q^2 + \beta$ gravity.}		
  		
  \end{figure}
  %%%%%%%%%%%%%%%%%%%%%%%%%%%%%%%%%%%%%%%%%%%%%% Figure 8 %%%%%%%%%%%%%%%%%%%%%%%%%%%%%%%%%%%%%%%%%%%%%%%%%%%
  \begin{figure}
  	(a)\includegraphics[width=7cm, height=6cm, angle=0]{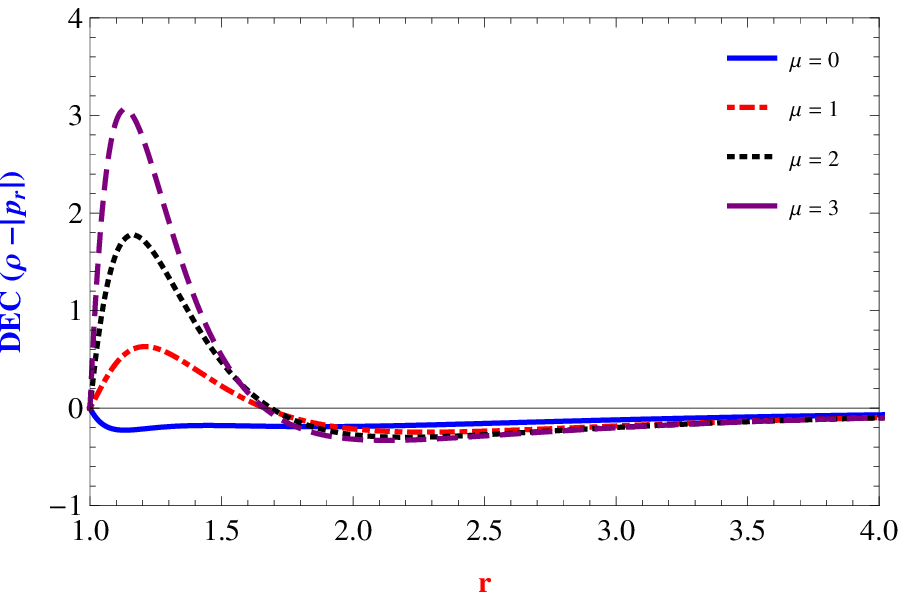}
  	(b)\includegraphics[width=7cm, height=6cm, angle=0]{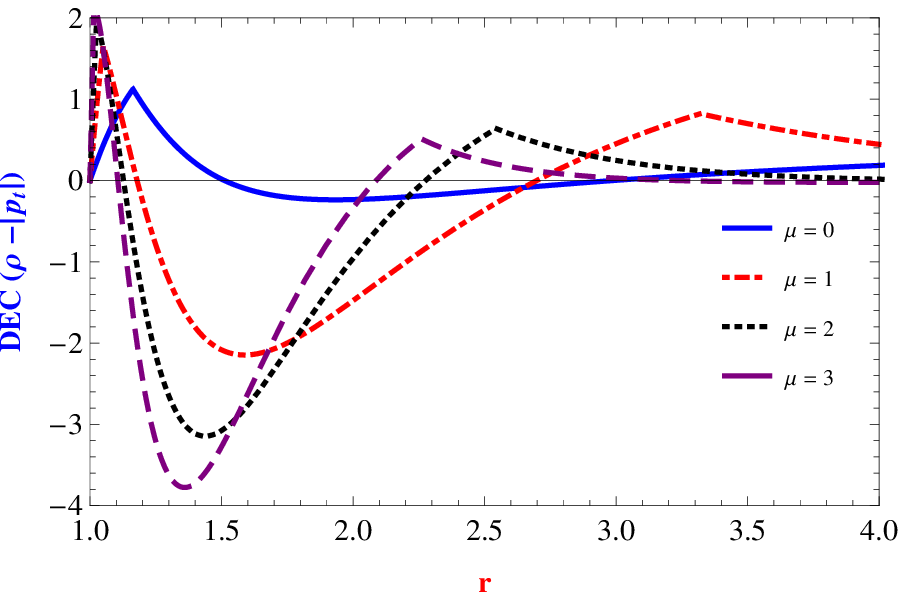}
  	\caption {Plots of radial  DEC ($\rho -|p_r|)$, and tangential DEC ($ \rho-|p_t|$) with throat radius $r_0 = 1$ in $f(Q)=\alpha Q^2 + \beta$ gravity.}
  \end{figure}
  
  %%%%%%%%%%%%%%%%%%%%%%%%%%%%%%%%%%%%%%%%%%%%%%%%%%%%%%%%%%%
 
 \subsection{quadratic Form  $f(Q)= \alpha Q^2+\beta Q + \gamma $}
 
  In this section, we take a more general quadratic form of function $f(Q)$ i.e. $f(Q)= \alpha Q^2+\beta Q + \gamma $. We examine this case for the possible TWH solution with non-exotic matter. Using this quadratic model with Yukawa-Casimir shape function, the field equations (\ref{eq12}), (\ref{eq13}) and (\ref{eq14}) can be developed to give the stress energy components as
 
 \begin{eqnarray}\label{eq26}
 \rho&=&\frac{1}{6 r^8 (3 r+{r_0})^2} \left\lbrace e^{-2 \mu  r} \left(-4 \alpha  {r_0}^2 \left(2 r^2+3 r {r_0}+{r_0}^2\right) e^{2 \mu  {r_0}} \left(12 (\mu +2) r^3+2 r^2 (5 (\mu +2) {r_0}-9)\right.\right.\right.\nonumber\\
& +&
\left. \left. \left.  r {r_0} (2 (\mu +2) {r_0}-21)-7 {r_0}^2\right)+3 r^4 e^{2 \mu  r} \left(9 \gamma  r^6+6 r^5 (6 \beta +\gamma  {r_0})+r^4 \left(\gamma  {r_0}^2+12 \beta  (2 {r_0}-3)\right)\right.\right.\right.\nonumber\\
&-&
\left. \left. \left. 4 r^3 (36 \alpha -\beta  ({r_0}-12) {r_0})-12 r^2 \left(\beta  {r_0}^2+\alpha  (16 {r_0}-9)\right)-24 \alpha  r {r_0} (2 {r_0}-9)+108 \alpha  {r_0}^2\right)\right.\right.\nonumber\\
&-&
 \left. \left. 2 r^2 {r_0} e^{\mu  (r+{r_0})} \left(18 \beta  (\mu +2) r^6+3 \beta  r^5 (7 (\mu +2) {r_0}-12)+r^4 (\beta  {r_0} (8 (\mu +2) {r_0}-57) \right.\right.\right.\nonumber\\
 &-&
  \left. \left.\left. 72 \alpha  (\mu +4))+r^3 \left(\beta  {r_0}^2 ((\mu +2) {r_0}-30)-12 \alpha  (11 (\mu +4) {r_0}-18)\right)-r^2 {r_0} \left(5 \beta  {r_0}^2 \right.\right.\right.\right.\nonumber\\
  &+& 
  \left. \left. \left.\left.  72 \alpha  ((\mu +4) {r_0}-7)\right)-12 \alpha  r {r_0}^2 ((\mu +4) {r_0}-32)+96 \alpha  {r_0}^3\right)\right)\right\rbrace 
 \end{eqnarray}
 
 \begin{eqnarray}\label{eq27}
 p_r& =&\frac{1}{2 r^8 (3 r+{r_0})^2}\left\lbrace e^{-2 \mu  r} \left(12 \alpha  {r_0}^2 \left(2 r^2+3 r {r_0}+{r_0}^2\right)^2  e^{2 \mu  {r_0}}+2 r^2 {r_0} \left(2 r^2+3 r {r_0}+{r_0}^2\right) \right.\right.\nonumber\\
 &\times&
 \left.\left. e^{\mu  (r+{r_0})} \left(3 \beta  r^3+\beta  r^2 {r_0}-24 \alpha  r-32 \alpha  {r_0}\right)+r^4 \left(-e^{2 \mu  r}\right) \left(9 \gamma  r^6+6 \gamma  r^5 {r_0}+\gamma  r^4 {r_0}^2 6 r^8 (3 r+{r_0})^3\right.\right.\right.\nonumber\\
 &+&
  \left. \left. \left. 12 \beta  r^3 {r_0}+4 r^2 \left(\beta  {r_0}^2-9 \alpha \right)-120 \alpha  r {r_0}-84 \alpha  {r_0}^2\right)\right)\right\rbrace
 \end{eqnarray}
 
 \begin{eqnarray}\label{eq28}
 p_t&=& \frac{1}{6 r^8 (3 r+{r_0})^3}\left\lbrace e^{2 \mu  ({r_0}-r)} \left(-3 r^4 e^{2 \mu  (r-{r_0})} \left(27 \gamma  r^7+27 r^6 (2 \beta +\gamma  {r_0})+9 r^5 {r_0} (8 \beta +\gamma  {r_0})\right.\right.\right.\nonumber\\
 &+&
  \left. \left.\left. r^4 \left({r_0} \left(\gamma  {r_0}^2+6 \beta  (5 {r_0}-3)\right)-216 \alpha \right)+4 r^3 \left(\beta  {r_0}^3-27 \alpha  (4 {r_0}+1)\right)+2 r^2 {r_0} \left(\beta  {r_0}^2\right.\right.\right.\right.\nonumber\\
  &-&
\left. \left. \left.\left. 6 \alpha  (22 {r_0}+15)\right)-12 \alpha  r {r_0}^2 (4 {r_0}+11)-60 \alpha  {r_0}^3\right)-4 \alpha  {r_0}^2 \left(2 r^2+3 r {r_0}+{r_0}^2\right) \left(18 \mu  r^9\right.\right.\right.\nonumber\\
&+&
 \left. \left.\left. 9 r^8 (3 \mu  {r_0}+2)+r^7 {r_0} (13 \mu  {r_0}+36)+2 r^6 {r_0}^2 (\mu  {r_0}+11)+4 r^5 {r_0}^3-36 r^4-18 r^3 (3 {r_0}+1)\right.\right.\right.\nonumber\\
 &-&
 \left. \left.\left. r^2 {r_0} (26 {r_0}+21)-4 r {r_0}^2 ({r_0}+4)-5 {r_0}^3\right)- r^2 {r_0} e^{\mu  (r-{r_0})} \left(54 \beta  \mu  r^{12}+9 \beta  r^{11} (11 \mu  {r_0}+6)\right.\right.\right.\nonumber\\
 &+&
 \left. \left.\left.  6 r^{10} \left(-36 \alpha  \mu +11 \beta  \mu  {r_0}^2+21 \beta  {r_0}\right)+r^9 \left(\beta  {r_0}^2 (19 \mu  {r_0}+102)-108 \alpha  (5 \mu  {r_0}+2)\right)\right.\right.\right.\nonumber\\
 &+&
\left. \left.\left. 2 r^8 {r_0} \left(\beta  {r_0}^2 (\mu  {r_0}+17)-12 \alpha  (20 \mu  {r_0}+27)\right)+4 r^7 \left(-27 \beta +\beta  {r_0}^4-45 \alpha  \mu  {r_0}^3-174 \alpha  {r_0}^2\right)\right.\right.\right.\nonumber\\
&-&
\left. \left.\left. 6 r^6 \left(4 \alpha  \mu  {r_0}^4+52 \alpha  {r_0}^3+33 \beta  {r_0}\right)-12 r^5 \left(4 \alpha  \left({r_0}^4-18\right)+\beta  {r_0} (11 {r_0}-3)\right)\right.\right.\right.\nonumber\\
&+&
\left. \left.\left. 2 r^4 \left(\beta  (9-19 {r_0}) {r_0}^2+216 \alpha  (5 {r_0}+1)\right)-4 r^3 {r_0} \left(\beta  {r_0}^2 ({r_0}+1)-6 \alpha  (80 {r_0}+39)\right)\right.\right.\right.\nonumber\\
&+&
\left. \left.\left. r^2 \left(-2 \beta  {r_0}^4+720 \alpha  {r_0}^3+888 \alpha  {r_0}^2\right)+24 \alpha  r {r_0}^3 (4 {r_0}+21)+120 \alpha  {r_0}^4\right)\right)\right\rbrace
 \end{eqnarray} 
 
   \begin{figure}
   	(a)\includegraphics[width=7cm, height=6cm, angle=0]{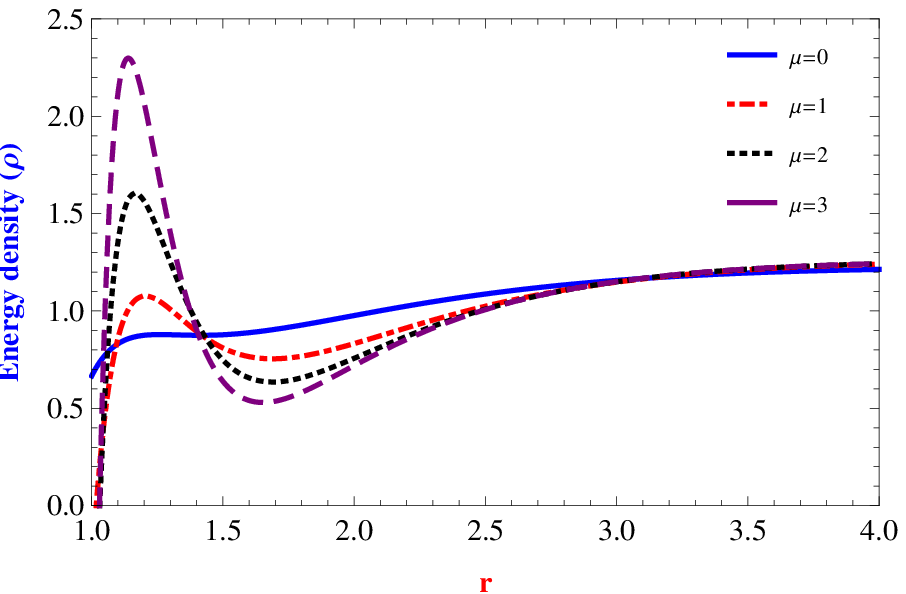}
   	(b)\includegraphics[width=7cm, height=6cm, angle=0]{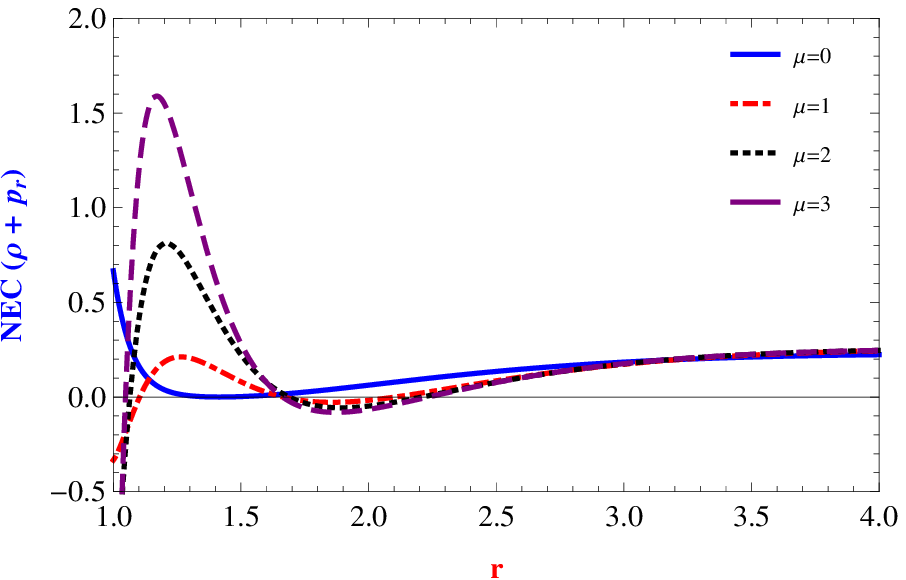}
   	\caption {Plots of energy density ($\rho$) and radial NEC ($\rho + p_r$) with throat radius $r_0 =1$ in $f(Q)=\alpha Q^2 + \beta Q +\gamma$ gravity.}
   \end{figure}
   %%%%%%%%%%%%%%%%%%%%%%%%%%%%%%%%% Figure 7 %%%%%%%%%%%%%%%%%%%%%%%%%%%%%%%%%%%%%%%%%%%%%%%%%%%%%%%%%%%%%%%%%%%%%%%%%%%%%%%
   \begin{figure}
   	(a)\includegraphics[width=7cm, height=6cm, angle=0]{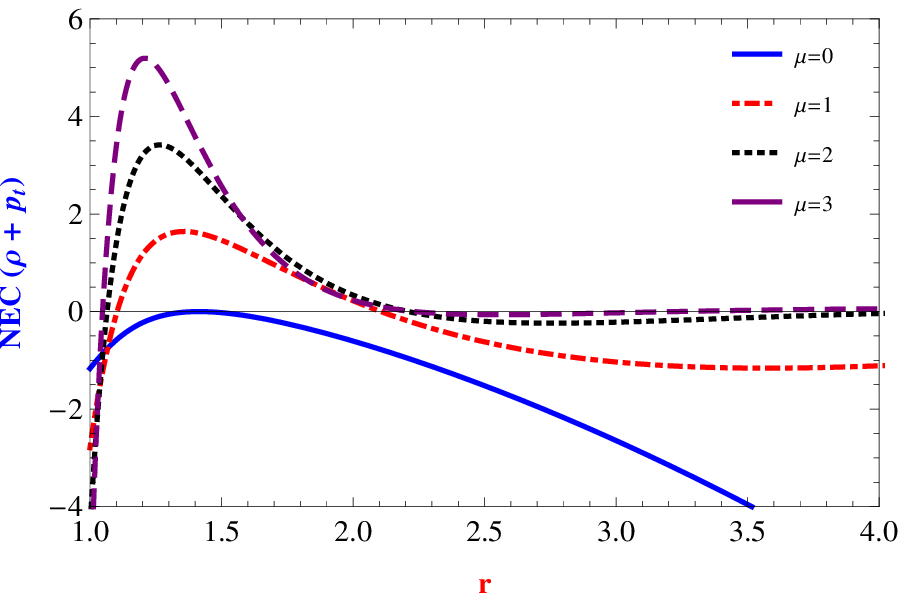}
   	(b)\includegraphics[width=7cm, height=6cm, angle=0]{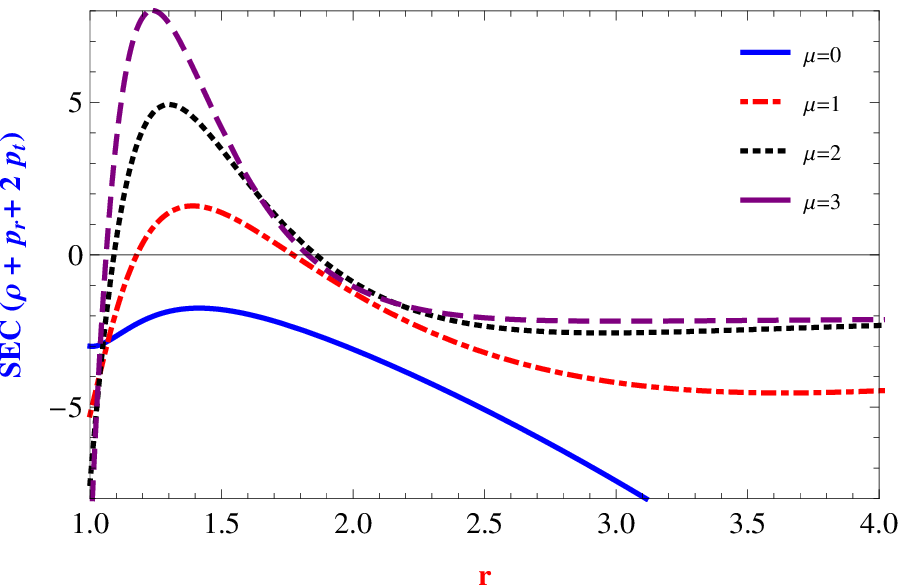}
   	\caption {Plots of tangential NEC ($\rho +p_t$) and SEC ($\rho + p_r+ 2p_t$) with throat radius $r_0=1$ in $f(Q)=\alpha Q^2 + \beta Q +\gamma$ gravity.}		
   	   \end{figure}
   %%%%%%%%%%%%%%%%%%%%%%%%%%%%%%%%%%%%%%%%%%%%%% Figure 8 %%%%%%%%%%%%%%%%%%%%%%%%%%%%%%%%%%%%%%%%%%%%%%%%%%%
   \begin{figure}
   	(a)\includegraphics[width=7cm, height=6cm, angle=0]{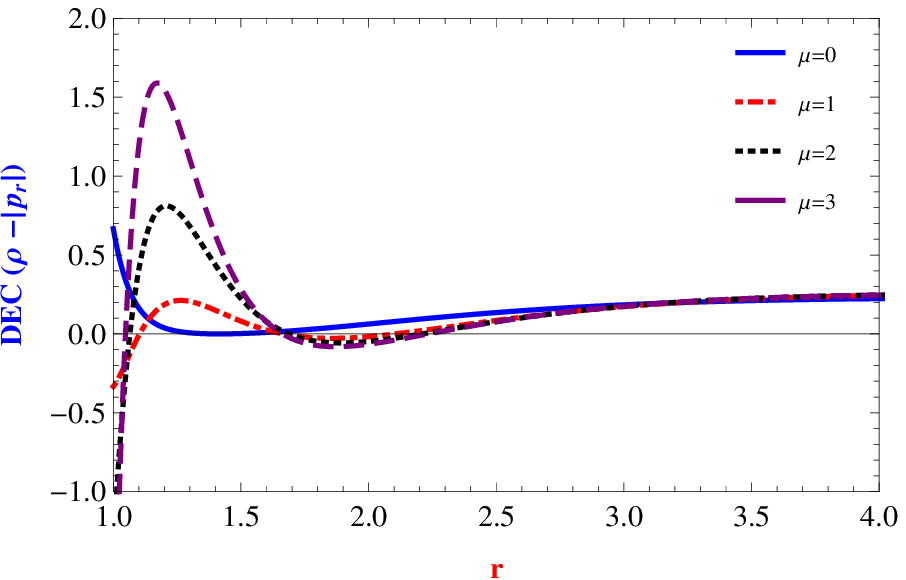}
   	(b)\includegraphics[width=7cm, height=6cm, angle=0]{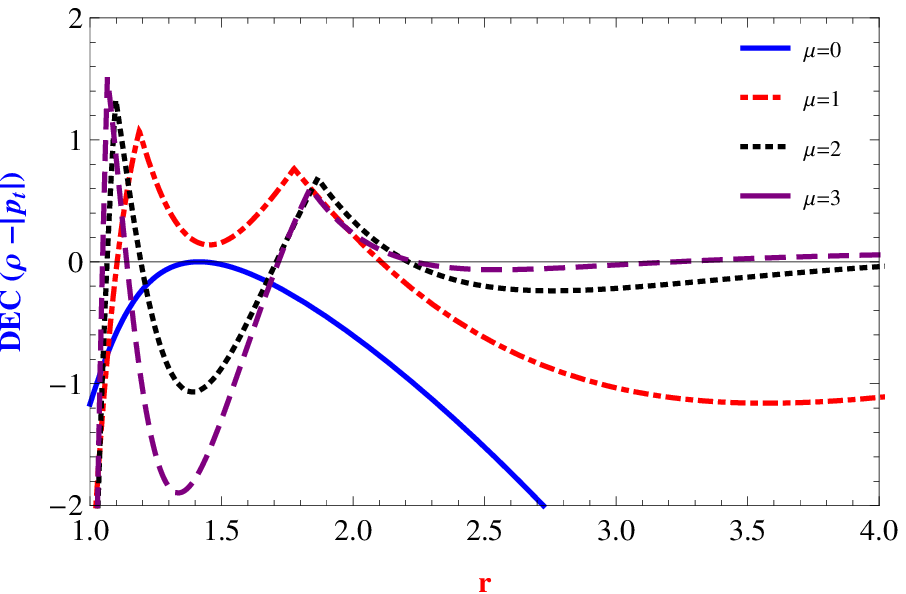}
   		\caption {Plots of radial  DEC ($\rho -|p_r|)$, and tangential DEC ($ \rho-|p_t|$) with throat radius $r_0 = 1$ in $f(Q)=\alpha Q^2 + \beta Q +\gamma$ gravity.}
   \end{figure}
%%%%%%%%%%%%%%%%%%%%%%%%%%%%%%%%%%%%%%%%%%%%%%%%%%%%%%%%%%%%%%%%%%%%%%%%%
%%%%%%%%%%%%%%%%%%%%%%%%%%%%%%%%%%%%%%%%%%%%%%%%%%%%%%%%%%%
%%%%%%%%%%%%%%%%%%%%%%%%%%%%%%%%%%%%%%%%%%%%%%%%%%%%%%%%%%%%
%%%%%%%%%%%%%%%%%%%%%%%%%%%%%%%%%%%%%%%%%%%%%%%%%%%%%%%%%%%%%%%%%%%%

\begin{figure}
	(a)\includegraphics[width=7cm, height=6cm, angle=0]{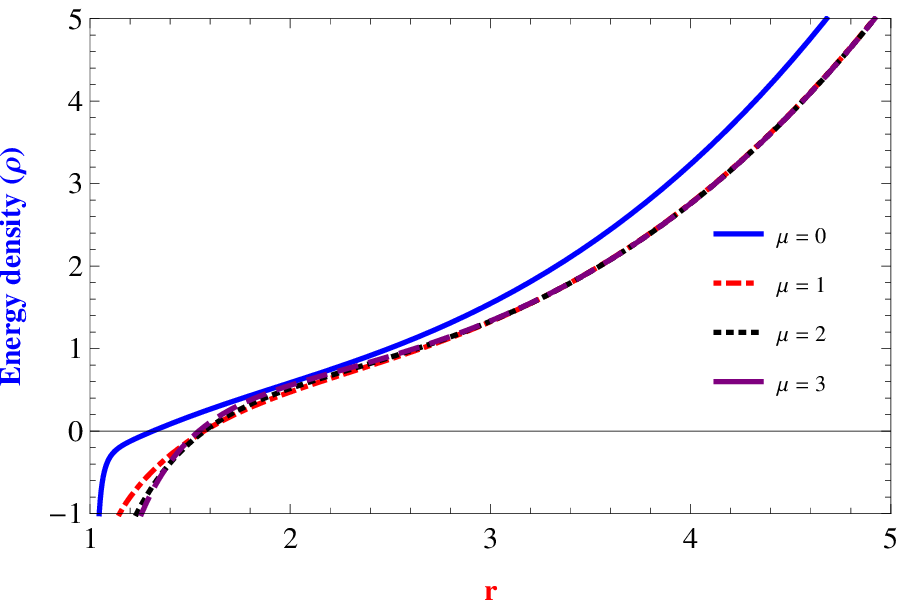}
	(b)\includegraphics[width=7cm, height=6cm, angle=0]{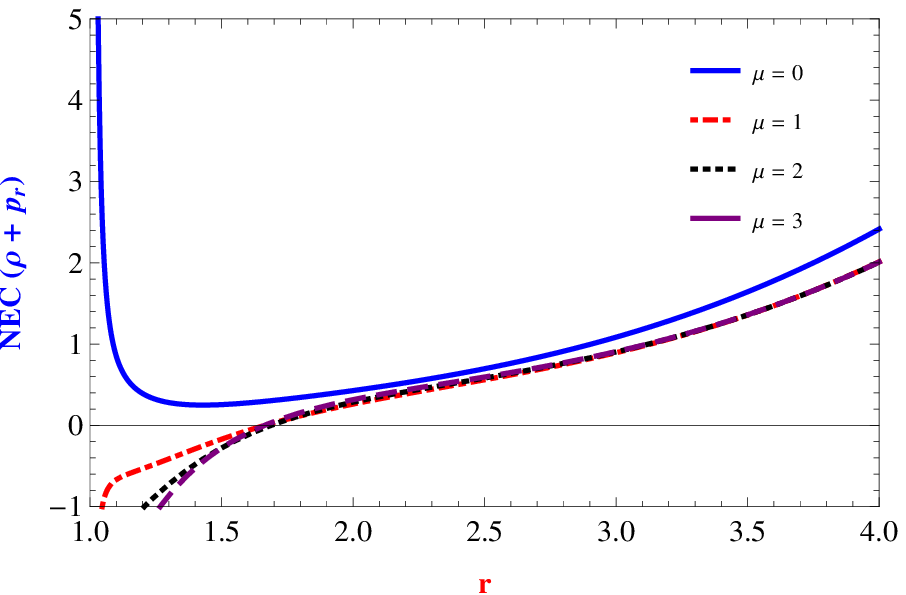}
	\caption {Plots of energy density ($\rho$) and radial NEC ($\rho + p_r$) with throat radius $r_0 =1$ in $f(Q)=  Q +\frac{\alpha}{Q}$ gravity. }
\end{figure}
%%%%%%%%%%%%%%%%%%%%%%%%%%%%%%%%% Figure 7 %%%%%%%%%%%%%%%%%%%%%%%%%%%%%%%%%%%%%%%%%%%%%%%%%%%%%%%%%%%%%%%%%%%%%%%%%%%%%%%
\begin{figure}
	(a)\includegraphics[width=7cm, height=6cm, angle=0]{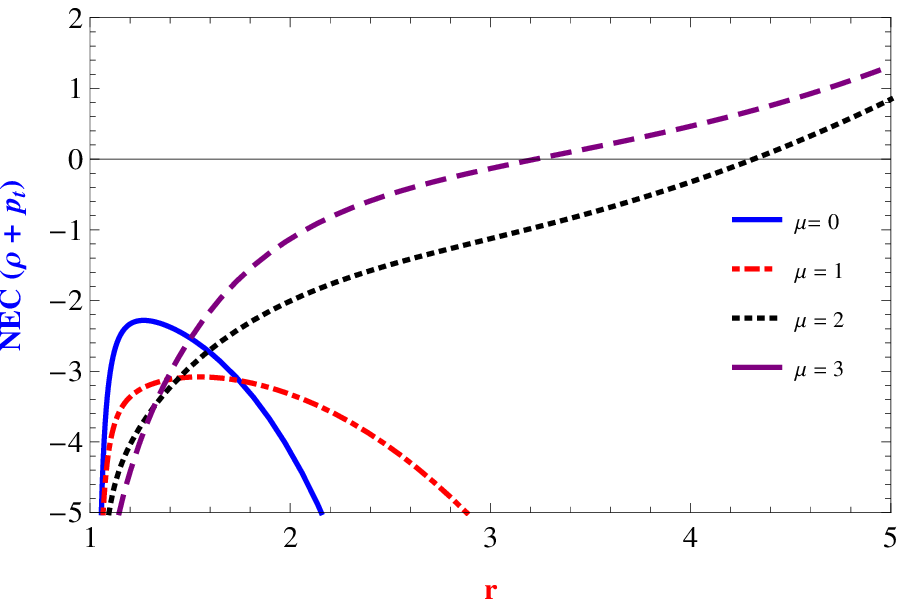}
	(b)\includegraphics[width=7cm, height=6cm, angle=0]{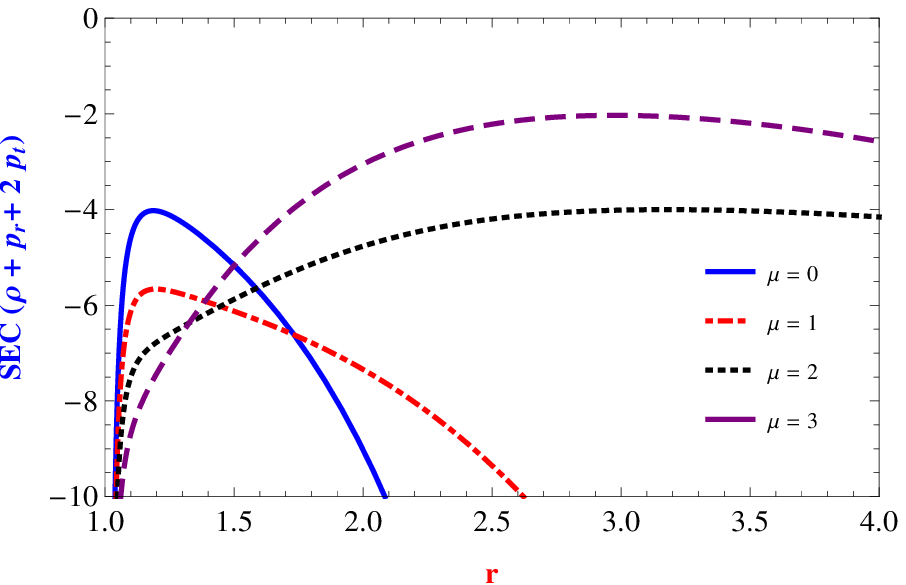}
	\caption {Plots of tangential NEC ($\rho +p_t$) and SEC ($\rho + p_r+ 2p_t$) with throat radius $r_0=1$ in $f(Q)=  Q +\frac{\alpha}{Q}$ gravity.}
\end{figure}
%%%%%%%%%%%%%%%%%%%%%%%%%%%%%%%%%%%%%%%%%%%%%% Figure 8 %%%%%%%%%%%%%%%%%%%%%%%%%%%%%%%%%%%%%%%%%%%%%%%%%%%
\begin{figure}
	(a)\includegraphics[width=7cm, height=6cm, angle=0]{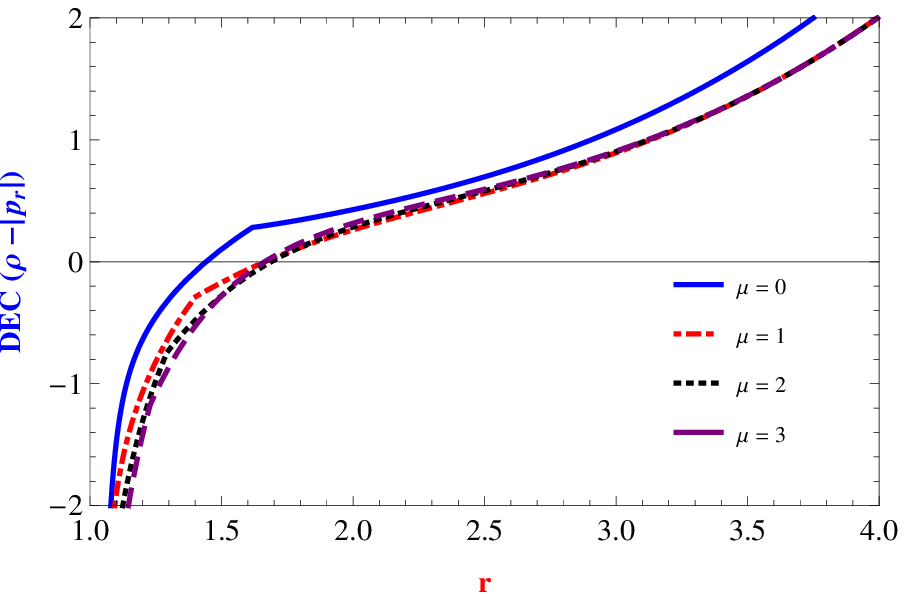}
	(b)\includegraphics[width=7cm, height=6cm, angle=0]{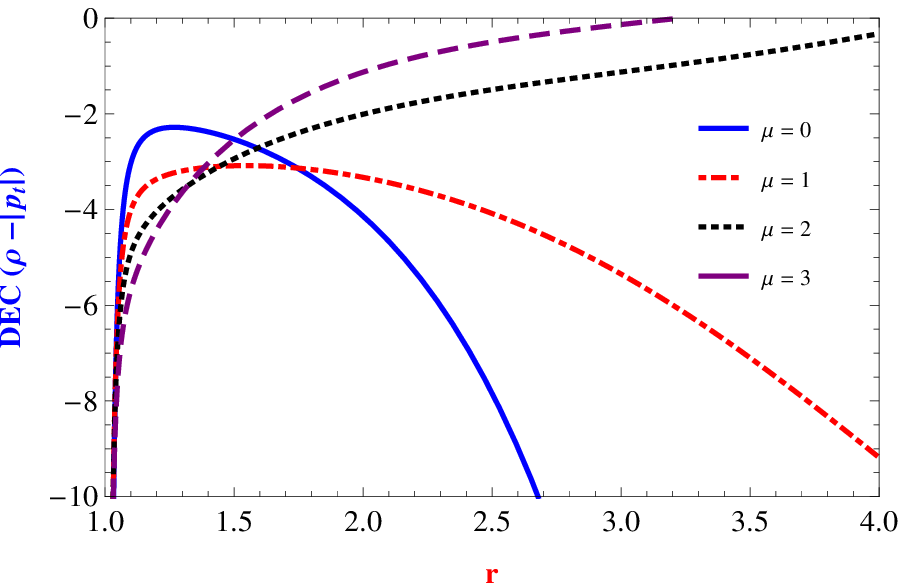}
	\caption {Plots of radial  DEC ($\rho -|p_r|)$, and tangential DEC ($ \rho-|p_t|$) with throat radius $r_0 = 1$ in $f(Q)=  Q +\frac{\alpha}{Q}$ gravity.}
\end{figure}

%%%%%%%%%%%%%%%%%%%%%%%%%%%%%%%%%%%%%%%%%%%%%%%%%%%%%%%%%  
   
 Fig. 7(a) depicts the energy density $\rho$ against all four values $0,1,2,3$ of scalar $\mu$. As we interpret from the figure, energy density is positive near the throat and throughout the region $r\geq r_0$. The NECs can be analyzed from the plots namely Figs. 7(b) and 8(a). It is very interesting to see that similar to the previous case, both the NECs are satisfied near the throat implying the presence of non-exotic matter near the wormhole throat except for the case of the original Casimir wormhole i.e. at $\mu=0$. For $\mu=0$ only the radial null energy condition is satisfied but tangential NEC is violated. Here, we also found the range of radial coordinates where NEC terms are satisfied. The radial NEC is satisfied for $r\in (1.04, 1.65)$ and tangential NEC is also satisfied in $r \in (1.04, 2.1)$ for Yukawa-Casimir energy but tangential NEC is violated everywhere in the region for Casimir source. The nature of SEC and both DECs are shown in Figs. 8(b), 9(a) and 9(b).

 %%%%%%%%%%%%%%%%%%%%%%%%%%%%%%%%%%%%%%%%%%%%%%%%%%%%%%%%%%%%

 \subsection{Inverse Power Law form:  $f(Q)=  Q+\frac{\alpha}{Q}$}
 
 This section is devoted to the results obtained on the stage of the inverse power law form of teleparallel gravity \cite{ref58}. The energy profile for stress-energy tensor in terms of radial pressure $p_r$, tangential pressure $p_t$ and the energy density $\rho$ is obtained from (\ref{eq12}), (\ref{eq13}), (\ref{eq14}) as

\begin{eqnarray}\label{eq29}
	\rho&=&\frac{1}{12 r^4 (r+{r_0})^2 (3 r+{r_0}) \left({r_0} (2 r+{r_0})-3 r^2 e^{\mu  (r-{r_0})}\right)^2}\left\lbrace e^{\mu  ({r_0}-r)} \left(24 r^2 {r_0}^2 (r+{r_0})^2 (2 r \right.\right.\nonumber\\
	&+&
	\left.\left. {r_0}) e^{\mu  (r-{r_0})} \left(6 (\mu +3) r^3+r^2 (5 (\mu +3) {r_0}-18)+r {r_0} ((\mu +3) {r_0}-24)-8 {r_0}^2\right)\right.\right.\nonumber\\
	&-&
	\left.\left. 4 {r_0}^3 \left(2 r^2+3 r {r_0}+{r_0}^2\right)^2 \left(6 (\mu +2) r^3+r^2 (5 (\mu +2) {r_0}-12)+r {r_0} ((\mu +2) {r_0}-15)-5 {r_0}^2\right)\right.\right.\nonumber\\
	&-&
	\left.\left. 6 r^6 \left(27 \alpha  r^8+27 \alpha  r^7 {r_0}+9 \alpha  r^6 {r_0}^2+\alpha  r^5 {r_0}^3-108 r^4-36 r^3 (7 {r_0}-3)-36 r^2 {r_0} (5 {r_0}-9)\right.\right.\right.\nonumber\\
	&-&
	\left. \left.\left. 36 r ({r_0}-9) {r_0}^2+108 {r_0}^3\right) e^{3 \mu  (r-{r_0})}+r^4 {r_0} e^{2 \mu  (r-{r_0})} \left(54 \alpha  (\mu +2) r^9+81 \alpha  (\mu +2) r^8 {r_0}\right.\right.\right.\nonumber\\
	&+&
	\left.\left.\left.  9 \alpha  r^7 {r_0} (5 (\mu +2) {r_0}+3)+\alpha  r^6 {r_0}^2 (11 (\mu +2) {r_0}+27)+r^5 \left(-216 (\mu +6)+\alpha  (\mu +2) {r_0}^4 \right.\right.\right.\right.\nonumber\\
	&+&
	\left. \left.\left.\left. 9 \alpha  {r_0}^3\right)+r^4 \left(\alpha  {r_0}^4-612 (\mu +6) {r_0}+1296\right)-36 r^3 {r_0} (17 (\mu +6) {r_0}-123)\right.\right.\right.\nonumber\\
	&-&
	\left. \left.\left. 36 r^2 {r_0}^2 (7 (\mu +6) {r_0}-155)-36 r {r_0}^3 ((\mu +6) {r_0}-85)+612 {r_0}^4\right)\right)\right\rbrace 
\end{eqnarray}
\begin{eqnarray}\label{eq30}
	p_r&=&\frac{1}{4 r^4 (r+{r_0})^2 (3 r+{r_0}) \left({r_0} (2 r+{r_0})-3 r^2 e^{\mu  (r-{r_0})}\right)^2}\left\lbrace  e^{\mu  ({r_0}-r)} \left(-8 r^2 {r_0}^2 (3 r+4 {r_0}) \left(2 r^2 \right.\right.\right.\nonumber\\
	&+&
	\left. \left.\left.  3 r {r_0}+{r_0}^2\right)^2 e^{\mu  (r-{r_0})}+4 {r_0}^3 \left(2 r^2+3 r {r_0}+{r_0}^2\right)^3-3 r^4 {r_0} \left(2 r^2+3 r {r_0}+{r_0}^2\right) \left(9 \alpha  r^6  \right.\right.\right.\nonumber\\
	&+&
	\left. \left.\left. 6 \alpha  r^5 {r_0}+\alpha  r^4 {r_0}^2-12 r^2-40 r {r_0}-28 {r_0}^2\right) e^{2 \mu  (r-{r_0})}+2 r^6 \left(27 \alpha  r^7+54 \alpha  r^6 {r_0}+27 \alpha  r^5 {r_0}^2\right.\right.\right.\nonumber\\
	&+&
	\left. \left.\left. 4 \alpha  r^4 {r_0}^3-36 r^2 {r_0}-72 r {r_0}^2-36 {r_0}^3\right) e^{3 \mu  (r-{r_0})}\right)\right\rbrace
\end{eqnarray}

\begin{eqnarray}\label{eq31} 
	p_t&=&\frac{1}{24 r^4 (r+{r_0})^2 (3 r+{r_0})^2 \left({r_0} (2 r+{r_0})-3 r^2 e^{\mu  (r-{r_0})}\right)^2}\left\lbrace e^{\mu  ({r_0}-r)} \left(24 r^2 {r_0}^2 (r+{r_0})^2 (2 r+{r_0})\right.\right.\nonumber\\
	&\times&
	\left. \left. e^{\mu  (r-{r_0})} \left(18 \mu  r^9+9 r^8 (3 \mu  {r_0}+2)+r^7 {r_0} (13 \mu  {r_0}+36)+2 r^6 {r_0}^2 (\mu  {r_0}+11)+4 r^5 {r_0}^3 \right.\right.\right.\nonumber\\
	&-&
	\left. \left.\left.  54 r^4-81 r^3 {r_0}+3 r^2 {r_0} (6-13 {r_0})+3 r {r_0}^2 (1-2 {r_0})-3 {r_0}^3\right)-4 {r_0}^3 \left(2 r^2+3 r {r_0}+{r_0}^2\right)^2 \right.\right.\nonumber\\ 
	&\times&
	\left. \left. \left(18 \mu  r^9+9 r^8 (3 \mu  {r_0}+2)+r^7 {r_0} (13 \mu  {r_0}+36)+2 r^6 {r_0}^2 (\mu  {r_0}+11)+4 r^5 {r_0}^3-36 r^4 \right.\right.\right.\nonumber\\
	&-&
	\left. \left.\left.  54 r^3 {r_0}+2 r^2 {r_0} (6-13 {r_0})+2 r {r_0}^2 (1-2 {r_0})-2 {r_0}^3\right)+6 r^6 \left(81 \alpha  r^9+27 \alpha  r^8 (5 {r_0}+6) \right.\right.\right.\nonumber\\
	&+&
	\left. \left.\left. 27 \alpha  r^7 {r_0} (3 {r_0}+11)+3 \alpha  r^6 {r_0}^2 (7 {r_0}+69)+r^5 \left(2 \alpha  {r_0}^4+63 \alpha  {r_0}^3-324\right)+r^4 {r_0} \left(7 \alpha  {r_0}^3 \right.\right.\right.\right.\nonumber\\
	&-&
	\left. \left.\left.\left.  972\right)-36 r^3 {r_0} (29 {r_0}-3)-36 r^2 {r_0}^2 (13 {r_0}-5)+36 r {r_0}^3 (1-2 {r_0})-36 {r_0}^4\right) e^{3 \mu  (r-{r_0})} \right.\right.\nonumber\\
	&+&
	\left. \left. r^4 {r_0} e^{2 \mu  (r-{r_0})} \left(162 \alpha  \mu  r^{15}+27 \alpha  r^{14} (13 \mu  {r_0}+6)+27 \alpha  r^{13} {r_0} (11 \mu  {r_0}+16)+3 \alpha  r^{12} {r_0}^2 (41 \mu  {r_0} \right.\right.\right.\nonumber\\
	&+&
	\left.\left.\left. 144)+r^{11} \left(-648 \mu +25 \alpha  \mu  {r_0}^4+204 \alpha  {r_0}^3\right)+2 r^{10} \left(\alpha  \left(\mu  {r_0}^5+23 {r_0}^4-162\right)-162 (7 \mu  {r_0} \right.\right.\right.\right.\nonumber\\
	&+&
	\left.\left.\left.\left. 2)\right)+r^9 \left(-648 \alpha +4 \alpha  {r_0}^5-3060 \mu  {r_0}^2-54 (13 \alpha +48) {r_0}\right)-18 r^8 {r_0} \left(84 \alpha +110 \mu  {r_0}^2 \right.\right.\right.\right.\nonumber\\
	&+&
	\left.\left.\left.\left. (33 \alpha +224) {r_0}\right)-6 r^7 {r_0}^2 \left(237 \alpha +102 \mu  {r_0}^2+(41 \alpha +504) {r_0}\right)-2 r^6 \left(36 \mu  {r_0}^5+5 (5 \alpha \right.\right.\right.\right.\nonumber\\
	&+&
	\left.\left.\left.\left. 108) {r_0}^4+333 \alpha  {r_0}^3-1944\right)-2 r^5 {r_0} \left(2 (\alpha +36) {r_0}^4+77 \alpha  {r_0}^3-6804\right)-2 r^4 {r_0} \left(7 \alpha  {r_0}^4 \right.\right.\right.\right.\nonumber\\
	&-&
	\left.\left.\left.\left. 9180 {r_0}+648\right)+216 r^3 {r_0}^2 (55 {r_0}-13)+216 r^2 {r_0}^3 (17 {r_0}-7)+216 r {r_0}^4 (2 {r_0}+1)\right.\right.\right.\nonumber\\
	&+&
	\left.\left.\left. 216 {r_0}^5\right)\right)\right\rbrace
\end{eqnarray}  

In power law form, various energy conditions and other profiles are coined in Fig. (10)- Fig. (12). The graphs are plotted against  $\mu = 0,1,2,3$. The original Casimir energy density is found positive in the region $r\in (1.3, \infty)$ and Yukawa-Casimir energy density is also observed positive for every non-zero value of $\mu$ in the region $r\in (1.55, \infty)$ which are plotted in Fig. 10(a). The radial NEC is validated everywhere in the region for Casimir energy and it is also validated in the region $r\in (1.55, \infty)$ for Yukawa-Casimir energy. In the case of Casimir energy the tangential NEC is violated for $\forall r\geq r_0$ in the region, but it is satisfied for $\mu= 3,4$ when radial coordinate $r\in (4.3, \infty)$ and $r\in (3.2, \infty)$ respectively. The SEC is violated in both the cases in Fig. 11(b) for all values of radial coordinate in the region. The radial DEC in Fig. 12(a) is validated in $r\in (1.45, \infty)$ for $\mu=0$ and $r\in (1.65, \infty)$ for $\mu= 1,2,3$. In Fig. 12 (b) the tangential DEC is violated in both cases.

\section{Discussion and Conclusion}
  
An unfortunate repercussion of GR regarding the traversability of the wormhole is the unavoidable breach of null energy conditions which leads to the presence of exotic or non-ordinary matter in the WH throat. The focus nowadays is on finding the solutions of traversable wormholes threaded by ordinary matter or with a minimal amount of exotic matter by using the modified gravity framework. The usual energy conditions are satisfied by the classical matter. Hence it is possible that wormholes must come from the domain of semi-classical or more likely a feasible quantum theory of the gravitational field. It's a well-known fact that Casimir energy is the only source of negative or exotic energy for various physical systems. In \cite{refa} author conjectured that the traversability could be supported by the quantum fluctuations as an effective source of the semiclassical Einstein equation.
In this present work, we aim to find the solution for the traversable wormhole with ordinary matter threading the wormhole throat. In this regard, we examine the static and spherically symmetric wormhole in the backdrop of symmetric teleparallel gravity where the non-metricity term $Q$ defines the gravitational interactions of space-time.\\

The motivation of this current study comes from \cite{ref58} where various forms of $f(Q)$ gravity were discussed for the plausible solutions of traversable WH. The outcomes in \cite{ref58} indicate the wormhole solutions supported by the exotic matter which gives the requisite negative energy to sustain the wormhole throat. As stated earlier in this section, the violation of NEC is accountable for the traversability which order indicates the existence of exotic matter in the wormhole throat. To circumvent this situation and to approach a more realistic solution supported by ordinary matter, we induce the Yukawa- Casimir wormhole apparatus in the backdrop of symmetric teleparallel gravity taking four function forms of $f(Q)$. The Yukawa modification of the Casimir wormhole is obtained where zero tidal force is imposed with the help of the equation of state \cite{refc}. The Casimir energy is the source of exotic energy which is negative energy and can support the traversability of the wormhole geometry. Introducing the Yukawa term in the original Casimir wormhole the Yukawa Casimir wormhole is constructed. We try to set up the wormhole geometry threaded by the non-exotic matter where the requisite negative energy is sourced from the Casimir energy instead of exotic matter \\

In the case of linear form $f(Q) = \alpha Q$, we gather from the plots that in all the cases of original Casimir and Yukawa-Casimir, the energy density $\rho$ is positive for $r\geq 1.6$.  We also found that wormhole throat is filled with exotic matter where NECs are violated in Casimir energy source and Yukawa-Casimir energy source for $\mu= 1,2$ but for $\mu =3$ NECs are satisfied in region $r \in (3.5, \infty) $. In the second attempt of power law form $f(Q)= \alpha Q^2 +\beta$, we observed that energy density $\rho$ is positive for all values of radial coordinate in both cases. We also found wormhole throat is filled with the non-exotic matter as both NECs are satisfied at the throat and in region $r\in (1, \infty)$ after Yukawa modification in Casimir energy. It is interesting to see that radial NEC is violated here in the original Casimir energy source. In the next case we have taken a more general quadratic form $f(Q)= \alpha Q^2 +\beta Q +\gamma$, we again found energy density positive, and Both NECs are satisfied in a specific region after Yukawa modification. Radial NEC is satisfied in $r\in (1.04, 1.65)$ and tangential NEC is also satisfied in $r\in (1.04, 2.1)$. We again analyzed the clear difference between Casimir energy and Yukawa-Casimir energy as tangential NEC is violated in Casimir energy for $\forall r\geq r_0$. In the last case we investigated inverse power law form $f(Q)= Q+\frac{\alpha}{Q}$, we observed that tangential NEC is violated in Casimir energy and Yukawa-Casimir for $\mu= 1$ but it is also satisfied for $\mu=2, 3$ in regions $r \in (4.3, \infty)$ and $r \in (3.2, \infty)$ respectively. Here, we again found a specific region in which both NECs are satisfied after Yukawa modification in Casimir energy.      
             
In this investigation we found for all specific forms of $f(Q)$ in Casimir source, null energy conditions are violated and wormholes are self-sustained traversable but after Yukawa modification in Casimir energy source violation of null energy conditions is restricted to some regions in the vicinity of the throat. These results are completely different from the previous wormhole solutions mentioned in \cite{ref58}, where authors investigated traversable wormholes violating null energy conditions in $f(Q)$ gravity.   Hence we may conclude that with semiclassical gravity theory, with appropriate choices the solution of traversable wormholes can be obtained where the traversability is achieved with the non-exotic matter. 
% 
%%%%%%%%%%%%%%%%%%%%%%%%%%%%%%%%%%%%%%%%%%%%%%%%%%%%%%%%%%%%%%%%%%%%%%%%%%%%%%%%%%%%%%%%%%%%%%%%%%%%%%5
%\section*{Acknowledgments} 

%%%%%%%%%%%%%%%%%%%%%%%%%%%%%%%%%%%%%%%%%%%%%%%%%%%%%%%%%%%%%%%%%%%%%%%%%%%%%%%%%%%%%%%%%%%%%%%%%%%%%%
%%%%%%%%%%%%%%%%%%%%%%%%%%%%%%%%%%%%%%%%%%%%%%%%%%%%%%%%%%%%%%%%%%%%%%%%%%%%%%%%%%%%%%%%%%%%%%%%%%%%%%%%%%%

\end{document}